\documentclass[twocolumn]{aastex631}
\usepackage{amsmath}

\received{31 January 2023}
\revised{27 June 2023}
\accepted{10 July 2023}
\acceptjournal{Astronomical Journal}

\shorttitle{The Dependence of IRMP Star Occurrence on Galactic Environment}
\shortauthors{Reeves et al.}

\begin{document}

\title{The Dependence of Iron-rich Metal-poor Star Occurrence on Galactic
Environment Supports an Origin in Thermonuclear Supernova Nucleosynthesis}

\correspondingauthor{Zachary Reeves}
\email{zackreeves96@gmail.com}

\author[0000-0002-0821-878X]{Zachary Reeves}
\affiliation{William H.\ Miller III Department of Physics and Astronomy,
Johns Hopkins University, 3400 N Charles St, Baltimore, MD 21218, USA}

\author[0000-0001-5761-6779]{Kevin C.\ Schlaufman}
\affiliation{William H.\ Miller III Department of Physics and Astronomy,
Johns Hopkins University, 3400 N Charles St, Baltimore, MD 21218, USA}

\author[0000-0001-6533-6179]{Henrique Reggiani}
\altaffiliation{Carnegie Fellow}
\affiliation{The Observatories of the Carnegie Institution for Science,
813 Santa Barbara St, Pasadena, CA 91101, USA}

\begin{abstract}

\noindent
It has been suggested that a class of chemically peculiar metal-poor stars
called iron-rich metal-poor (IRMP) stars formed from molecular cores with
metal contents dominated by thermonuclear supernova nucleosynthesis.
If this interpretation is accurate, then IRMP stars should be more
common in environments where thermonuclear supernovae were important
contributors to chemical evolution.  Conversely, IRMP stars should
be less common in environments where thermonuclear supernovae were not
important contributors to chemical evolution.  At constant $[\text{Fe/H}]
\lesssim -1$, the Milky Way's satellite classical dwarf spheroidal (dSph)
galaxies and the Magellanic Clouds have lower $[\text{$\alpha$/Fe}]$ than
the Milky Way field and globular cluster populations.  This difference
is thought to demonstrate the importance of thermonuclear supernova
nucleosynthesis for the chemical evolution of the Milky Way's satellite
classical dSph galaxies and the Magellanic Clouds.  We use data from
the Sloan Digital Sky Survey (SDSS) Apache Point Observatory Galactic
Evolution Experiment (APOGEE) and Gaia to infer the occurrence of
IRMP stars in the Milky Way's satellite classical dSph galaxies
$\eta_{\text{dSph}}$ and the Magellanic Clouds $\eta_{\text{Mag}}$
as well as in the Milky Way field $\eta_{\text{MWF}}$ and globular
cluster populations $\eta_{\text{MWGC}}$.  In order of decreasing
occurrence, we find $\eta_{\text{dSph}}=0.07_{-0.02}^{+0.02}$,
$\eta_{\text{Mag}}=0.037_{-0.006}^{+0.007}$,
$\eta_{\text{MWF}}=0.0013_{-0.0005}^{+0.0006}$, and a 1-$\sigma$
upper limit $\eta_{\text{MWGC}}<0.00057$.  These occurrences support
the inference that IRMP stars formed in environments dominated by
thermonuclear supernova nucleosynthesis and that the time lag between
the formation of the first and second stellar generations in globular
clusters was longer than the thermonuclear supernova delay time.

\end{abstract}


\section{Introduction}\label{intro}

Thermonuclear supernovae\footnote{In this article we use the phrase
``thermonuclear supernovae'' to refer to the theoretical concept of
electron-degenerate carbon--oxygen white dwarfs experiencing runway
carbon fusion that releases enough energy to gravitationally unbind the
white dwarfs and thereby cause explosions.} are prolific producers of
iron-peak elements with a theoretically predicted average stable yield of
more than $0.6~M_{\odot}$ of iron-peak elements but much less $\alpha$
and light odd-$Z$ elements.  This is in sharp contrast to core-collapse
supernovae that produce $\alpha$ and light odd-$Z$ elements in roughly
the solar ratios but with relatively little iron-peak production
\citep[e.g.,][]{suk16}.

Based on a compilation of theoretical stable nucleosynthetic yields
predicted by a variety of thermonuclear supernova progenitor channels
(i.e., single degenerate and double degenerate) and explosion mechanisms
(e.g., pure detonations, pure deflagrations, delayed detonations,
double detonations, etc.), \citet{reg23} proposed the existence of a
new class of chemically peculiar metal-poor stars with $[\text{Fe/H}]
\lesssim -1$ and $[\text{O,F,Ne,Na,Mg,Al,Cl,K,Co,Cu,Zn/Fe}] < 0$ formed
from molecular cores with metal contents dominated by thermonuclear
supernova nucleosynthesis.\footnote{The yields in that compilation came
from \citet{sei13,sei16}, \citet{fin14}, \citet{ohl14}, \citet{pap16},
\citet{leu18,leu20a,leu20b}, \citet{nom18}, \citet{bra19}, \citet{boo21},
\citet{gro21a,gro21b}, and \citet{neo22}.}  They called stars with
these properties iron-rich metal-poor (IRMP) stars, as this part of
elemental abundance space is consistent with thermonuclear supernova
nucleosynthesis but rarely observed in metal-poor stars.  They argued
that if their interpretation is correct, then IRMP stars should be more
common in environments where thermonuclear supernovae were relatively more
important contributors to chemical evolution relative to core-collapse
supernovae (e.g., environments with long star formation durations).
On the other hand, they argued that in environments where thermonuclear
supernovae were not important contributors to chemical evolution relative
to core-collapse supernovae (e.g., environments with short star formation
durations) IRMP stars should be less common.

One way to test this prediction would be to compare the relative
occurrence of IRMP stars in the Milky Way's field population, its
globular clusters, its satellite classical dwarf spheroidal (dSph)
galaxies, and the Magellanic Clouds.  At constant spectroscopically
inferred metallicities $[\text{Fe/H}] \lesssim -1$\footnote{In this
article metallicity $[\text{Fe/H}]$ has its usual meaning $[\text{Fe/H}] =
\log_{10}\left(N_{\text{Fe}}/N_{\text{H}}\right)_{\ast}-\log_{10}\left(N_{\text{Fe}}/N_{\text{H}}\right)_{\odot}$
where $N_{\text{X}}$ are the logarithmic number densities
of atoms of an element X in a stellar photosphere and
$N_{\text{H}} \equiv 12$.}, both the Milky Way's classical dSph
satellites \citep{she01,she03,tol03,kir09,kir10,kir11a,kir11b}
and the Magellanic Clouds \citep{pom08,van13,nid20} have
lower spectroscopically inferred ratios of the $\alpha$
elements oxygen, magnesium, silicon, and calcium to iron
$[\text{$\alpha$/Fe}]$\footnote{In this article the $\alpha$ element-to-iron
ratio $[\text{$\alpha$/Fe}]$ has its usual meaning $[\text{$\alpha$/Fe}] =
\log_{10}\left(N_{\alpha}/N_{\text{Fe}}\right)_{\ast}-\log_{10}\left(N_{\alpha}/N_{\text{Fe}}\right)_{\odot}$
where $\alpha$ is the sum of some subset of the elements
oxygen, magnesium, silicon, and calcium.} than the Milky Way
\citep[e.g.,][]{wal62,luc81,luc85,pet81,gra83,mag85,mag89,gra86,gra88,rya91,mcw95a,mcw95b}
and its globular clusters \citep[e.g.,][]{coh78,coh79,coh80,coh81,pil83}.
These offsets in $[\text{$\alpha$/Fe}]$ between the Milky Way field
\& globular cluster populations and its satellite classical dSph
galaxies \& the Magellanic Clouds are usually understood to indicate
the longer durations of low-metallicity star formation and therefore the
relatively more important contributions of thermonuclear supernovae to
the chemical evolution of the latter two environments at low metallicities
\citep[e.g.,][]{lan03,lan04,lan07,lan06,lan08,mar06,kir11a,bek12,nid20,reg21}.
If the \citet{reg23} interpretation of IRMP stars is valid, then
IRMP stars should be less common in the Milky Way field \& globular
cluster populations than in its satellite classical dSph galaxies \&
the Magellanic Clouds.

The nitrogen, sodium, and aluminum abundances of individual stars
in globular clusters have been found to be anticorrelated with
the carbon, oxygen, and magnesium abundances in the same stars.
These abundance anticorrelations have been interpreted as evidence
for multiple generations of star formation in globular clusters
\citep[e.g.,][]{sne92,gra01,gra04,car09,bas18}.  The \citet{reg23}
interpretation of IRMP stars also suggests the possibility that the
occurrence of IRMP stars in globular clusters can be used to constrain
the time lag between the formation of a globular cluster's first
and second stellar generations.  For Type Ia supernovae\footnote{In
this article we use the phrase ``Type Ia supernovae'' to refer to the
electromagnetic transients empirically classified as Type Ia supernovae
based on their observed properties.} delay times $\tau_{\text{Ia}}
\lesssim 100$ Myr, the Type Ia supernova rate $\Phi_{\text{Ia}}$ has
the value $\Phi_{\text{Ia}}\sim 10^{-12}$ yr$^{-1}$ $M_{\odot}^{-1}$
\citep[e.g.,][]{tot08,mao11,mao12,mao14,gra14}.  Assuming that rate and a
typical first generation initial globular cluster mass $M_{\text{MWGC}}
\sim 10^{6}~M_{\odot}$ \citep[e.g.,][]{con12}, order 10 Type Ia
supernovae should occur in a newly formed globular cluster every 10 Myr
after the Type Ia delay time has elapsed.  While the occurrence of the
short-period binaries necessary to produce most of the theoretically
predicted thermonuclear supernova progenitor systems is a factor of
about three lower in globular clusters' first generations than in the
field \citep{car03,luc15}, after the Type Ia supernovae delay time has
elapsed a few thermonuclear explosions should occur every 10 Myr during
the formation of a globular cluster.  The \citet{reg23} interpretation of
IRMP stars therefore implies that if the time lag between the formation
of first and second generation stars in globular clusters was shorter
than the typical thermonuclear supernova delay time, then the occurrence
of IRMP stars in globular clusters should be significantly lower than
in the Milky Way field.  If the the time lag between the formation of
first and second generation stars in globular clusters is longer than the
typical thermonuclear supernova delay time, then the occurrence of IRMP
stars in globular clusters and the Milky Way field should be comparable.

We argue that if the \citet{reg23} interpretation of the origin of
IRMP stars is accurate, then the occurrence of IRMP stars in the
Milky Way field population $\eta_{\text{MWF}}$, the Magellanic Clouds
$\eta_{\text{Mag}}$, and the Milky Way's satellite classical dSph
galaxies $\eta_{\text{dSph}}$ should be ordered $\eta_{\text{MWF}}
< \eta_{\text{Mag}} \sim \eta_{\text{dSph}}$.  If the time lag
between the formation of the first and second generations in globular
clusters was shorter than the thermonuclear supernova delay time,
then the occurrence of IRMP stars in the Milky Way globular cluster
population $\eta_{\text{MWGC}}$ should be $\eta_{\text{MWGC}} \lesssim
\eta_{\text{MWF}} < \eta_{\text{Mag}} \sim \eta_{\text{dSph}}$.  If the
time lag between the formation of the first and second generations in
globular clusters was longer than the thermonuclear supernova delay time,
then $\eta_{\text{MWGC}} \sim \eta_{\text{MWF}} < \eta_{\text{Mag}}
\sim \eta_{\text{dSph}}$.  In this article, we calculate the occurrence
of IRMP stars in the Milky Way's satellite classical dSph galaxies,
the Magellanic Clouds, the Milky Way field population, and in Milky Way
globular clusters.  We describe in Section \ref{data} the assembly of our
analysis samples and quantify in Section \ref{analysis} the occurrence
of IRMP stars in each environment.  We review the implications of those
occurrences in Section \ref{discussion} and conclude by summarizing our
findings in Section \ref{conclusion}.

\section{Data}\label{data}

To calculate the occurrence of IRMP stars in the Milky Way's satellite
classical dSph galaxies, the Magellanic Clouds, the Milky Way field
population, and in Milky Way globular clusters we use data derived
from spectra that were gathered during the third and fourth phases
of the Sloan Digital Sky Survey \citep[SDSS;][]{eis11,bla17} as
part of its Apache Point Observatory Galactic Evolution Experiment
\cite[APOGEE;][]{maj17}.  These spectra were collected with the
APOGEE spectrographs \citep{zas13,zas17,wil19,bea21,san21} on the
New Mexico State University 1-m Telescope \citep{hol10}, the Sloan
Foundation 2.5-m Telescope \citep{gun06}, and the 2.5-m Ir\'{e}n\'{e}e
du Pont Telescope \citep{bow73}.  As part of SDSS Data Release
(DR) 17 \citep{abd22}, these spectra were reduced and analyzed
with the APOGEE Stellar Parameter and Chemical Abundance Pipeline
\citep[ASPCAP;][]{all06,hol15,nid15,gar16} using an $H$-band line list,
MARCS model atmospheres, and model-fitting tools optimized for the APOGEE
effort \citep{alv98,gus08,hub11,ple12,smi13,smi21,cun15,she15,jon20}.

We use the CasJobs
portal\footnote{\url{http://skyserver.sdss.org/casjobs/}} and the
query described in the Appendix to generate our initial sample of
photospheric stellar parameters and elemental abundances for giant stars
with $\log{g} < 3.8$.  As described in the Appendix, we use a carefully
curated set of data quality flags to ensure the accuracy and precision
of those photospheric stellar parameters and elemental abundances.
We set to \texttt{null} any elemental abundance/elemental abundance
uncertainty pair that does not pass the data quality checks described
in the Appendix.  Corrections for departures from local thermodynamic
equilibrium are usually small for $H$-band elemental abundance inferences
\citep[e.g.,][]{oso20}, and we choose not to apply them in our analysis.

Following \citet{reg23}, we define an iron-rich metal-poor star as
a star with $[\text{Fe/H}] < -1$ and $[\text{O,Na,Mg,Al,K,Co/Fe}]
< 0$.  Those elemental abundance ratios form the intersection of the
IRMP criteria defined in \citet{reg23} and the list of elemental
abundances reliably inferred for giant stars as part of APOGEE
DR17\footnote{\url{https://www.sdss.org/dr17/irspec/abundances/}}.
For the purposes of the occurrence calculation described in the next
section, an IRMP star with $[\text{Fe/H}] < -1$ must have at least one
non-\texttt{null} abundance ratio $[\text{O/Fe}]$, $[\text{Na/Fe}]$,
$[\text{Mg/Fe}]$, $[\text{Al/Fe}]$, $[\text{K/Fe}]$, or $[\text{Co/Fe}]$
and all abundance ratios $[\text{O,Na,Mg,Al,K,Co/Fe}]$ either sub-solar
or \texttt{null}.

To accurately label stars with their correct galactic environments,
we first join the APOGEE data described above with data from Gaia
DR2 \citep{gaia16,gaia18,are18,eva18,lin18,lur18,rie18} and DR3
\citep{gaia16,gaia21,fab21,lin21a,lin21b,rie21,row21,tor21}.
We use the \texttt{apogee\_id} string to identify the
corresponding Gaia DR2 and DR3 \texttt{source\_id} long integers
by joining with the \texttt{gaiadr2.tmass\_best\_neighbour} and
\texttt{gaiadr3.tmass\_psc\_xsc\_best\_neighbour} tables available in the
Gaia archive \citep{sal17,mar19}.  Occasionally multiple Gaia DR2 and DR3
\texttt{source\_id} long integers are matched to the same object in the
2MASS Point Source Catalog \citep{skr06}.  In those cases, we associate
a 2MASS object with the closest Gaia DR2 and DR3 object that has (1) an
absolute 2MASS $K_{\text{s}}$-band magnitude $M_{K} < 2.31$ assuming Gaia
DR2- or DR3-prior informed geometric distances \citep{bai18,bai21} and
(2) Gaia--2MASS colors $0.5 < G-J < 2.3$, $0.6 < G-H < 3.3$, and $0.6 <
G-K_{\text{s}} < 3.4$ predicted by the MESA Isochrones \& Stellar Tracks
(MIST) grid for metal-poor giants in the range $-2.5 < [\text{Fe/H}] <
-1.0$ \citep{pax11,pax13,pax15,pax18,dot16,cho16}.

We then use the Gaia DR2 \texttt{source\_id} to identify Milky Way
satellite classical dSph galaxies or Magellanic Clouds members using the
Gaia DR2-based membership lists published in \citet{gaia18}.  We identify
globular cluster members using the Gaia DR3 \texttt{source\_id} and the
\citet{vas21} lists of stars with globular cluster membership probability
greater than 0.5, and this procedure results in a sample with at least one
star from 41 globular clusters over the metallicity range $-2.3 \lesssim
[\text{Fe/H}] \lesssim -1.0$.

Most stars observed as part SDSS-III/APOGEE and SDSS-IV/APOGEE-2 were
selected for observation by a procedure that sought to minimize age and
metallicity biases \citep[e.g.,][]{zas13,zas17}, and we use the targeting
flag \texttt{extratarg} = 0 in the table \texttt{apogeeStar} to select
those stars for our Milky Way field sample.  To ensure the cleanest Milky
Way field sample possible, we then remove from this sample any stars that
are identified as dSph, Magellanic Cloud, or globular cluster members
in \citet{gaia18} or \citet{vas21}.  Because the vast majority of stars
with $[\text{Fe/H}] \lesssim -1$ observed as part of SDSS-III/APOGEE and
SDSS-IV/APOGEE-2 are on halo-like orbits \citep[e.g.,][]{hay18}, our Milky
Way field sample can be thought of as a Milky Way halo sample.  We list
in Table \ref{tab01} our entire analysis sample including IRMP status
and galactic environment.  We report in Table \ref{tab02} the number of
stars classified as IRMP stars in each environment $N_{\ast,\text{IRMP}}$
along with the number of stars in our analysis sample in each environment
$N_{\ast,\text{tot}}$.

\begin{deluxetable*}{ccccc}
\tablecaption{Analysis Sample\label{tab01}}
\tablewidth{0pt}
\tablehead{
\colhead{APOGEE ID} &
\colhead{Gaia DR3 \texttt{source\_id}} &
\colhead{Gaia DR2 \texttt{source\_id}} &
\colhead{IRMP} &
\colhead{Environment}
}
\startdata
\hline
2M17165079$-$2422565 & 4111066558908989184   & 4111066558908989184   & False & Milky Way\\
2M17150296$-$2423503 & 4114045307692005632   & 4114045307692005632   & False & Milky Way\\
2M19154424$-$0604209 & 4211181280154376320   & 4211181280154376320   & False & Milky Way\\
2M19033822$+$1745138 & 4514220364269464832   & 4514220364269464832   & False & Milky Way\\
2M16574858$-$2156135 & 4126283868512500864    & 4126283868512500864    & False & Milky Way\\
2M18501947$+$2948368 & 2041393317228382848   & 2041393317228382848   & False & Milky Way\\
2M18132084$+$0112054 & 4275831399934633984    & 4275831399934633984  & False & Milky Way\\
2M17571005$-$3020262 & 4056215664653907456    & 4056215664653907456  & False & Milky Way\\
2M17415271$-$2715374 & 4060889448072712832   & 4060889448072712832 & False & Milky Way\\
2M19084424$-$0618527 & 4205916204326948736 & 4205916204326948736 & False & Milky Way
\enddata
\tablecomments{This table is published in its entirety in the
machine-readable format.  A portion is shown here for guidance regarding
its form and content.}
\end{deluxetable*}

\begin{deluxetable*}{lccc}
\tablecaption{Occurrence of IRMP Stars as a Function of
Environment\label{tab02}}
\tablewidth{0pt}
\tablehead{
\colhead{Galactic Environment} &
\colhead{$N_{\ast,\text{IRMP}}$} &
\colhead{$N_{\ast,\text{tot}}$} &
\colhead{Occurrence}
}
\startdata
Milky Way & 5 & 4247 & $0.0013_{-0.0005}^{+0.0006}$\\
\textbf{Globular Cluster Sum} & 0 & 1998 & $<0.00057$\\
\hline
LMC & 4 & 203 & $0.023_{-0.009}^{+0.012}$\\
SMC & 24 & 572 & $0.043_{-0.008}^{+0.009}$\\
\textbf{Magellanic Clouds Sum} & 28 & 775 & $0.037_{-0.006}^{+0.007}$\\
\hline
Sagittarius & 0 & 27 & $<0.04$\\
Ursa Minor & 0 & 8 & $<0.12$\\
Sextans & 1 & 6 & $0.2_{-0.1}^{+0.2}$\\
Sculptor & 8 & 70 & $0.12_{-0.03}^{+0.04}$\\
Draco & 0 & 9 & $<0.11$\\
Carina & 1 & 31 & $0.05_{-0.03}^{+0.05}$\\
\textbf{dSph Galaxies Sum} & 10 & 151 & $0.07_{-0.02}^{+0.02}$
\enddata
\end{deluxetable*}

The typical elemental abundance inference uncertainties in our analysis
sample are (0.04, 0.23, 0.03, 0.04, 0.08, 0.16) dex for ($[\text{O/Fe}]$,
$[\text{Na/Fe}]$, $[\text{Mg/Fe}]$, $[\text{Al/Fe}]$, $[\text{K/Fe}]$,
$[\text{Co/Fe}]$).  In any case, elemental abundance inference
uncertainties are irrelevant for the occurrence analyses presented in
Section \ref{analysis} if (1) the uncertainty distributions for each
elemental abundance inference for each individual star are symmetric and
(2) the uncertainty distributions have statistically indistinguishable
widths across all of the galactic environments we explored.
The individual elemental abundance inference uncertainties presented
in the SDSS DR17 version of the table \texttt{aspcapStar} are reported
as symmetric.  Additionally, we confirmed that the individual elemental
abundance uncertainty distributions for oxygen, sodium, magnesium,
aluminum, potassium, and cobalt have statistically indistinguishable
widths across all of the galactic environments we explored.  Both of
the conditions listed above are therefore met in our analysis sample.
Likewise, we argue that our procedure to handle data quality issues will
not bias the occurrence analyses we present in Section \ref{analysis}.
The reason is that \texttt{null} values impact less than about 0.2\% of
the magnesium abundance inferences in our sample.  Because we require
all non-\texttt{null} elemental inferences to meet our IRMP criteria,
we are able to exclude essentially all non-IRMP stars from our IRMP
sample using magnesium alone almost regardless of data quality issues.

The recent discovery of a very metal-poor star with elemental abundances
best explained by the nucleosynthesis expected in a pair-instability
supernova has focused attention on that explanation for stars with
significantly subsolar $[\text{Na/Fe}]$ and $[\text{$\alpha$/Fe}]$
abundance ratios \citep{xin23}.  While the stars in our analysis sample
have subsolar $[\text{Na/Fe}]$ and the $\alpha$-element abundance ratios
$[\text{O/Fe}]$ and $[\text{Mg/Fe}]$, none of the stars in our sample
have the strong odd--even abundance ratios predicted to be produced by
pair-instability supernovae \citep[e.g.,][]{heg02}.  We are therefore
confident that the IRMP stars we identify are related to thermonuclear
supernovae.

\section{Analysis}\label{analysis}

We model the number of IRMP stars $N_{\ast,\text{IRMP}}$ in a sample
of $N_{\ast,\text{tot}}$ candidates using a binomial distribution.
Following \citet{sch14} we exploit the fact that a Beta$(\alpha$,$\beta$)
distribution is a conjugate prior to the binomial distribution and will
result in a Beta distribution posterior for the occurrence of IRMP stars
in a sample.  Bayes's Theorem guarantees
\begin{eqnarray}
f(\theta|\mathbf{y}) & = & \frac{f(\mathbf{y}|\theta)f(\theta)}
	                        {\int f(\mathbf{y}|\theta)f(\theta)d\theta},
\end{eqnarray}
where $f(\theta|\mathbf{y})$ is the posterior distribution of the model
parameter $\theta$, $f(\mathbf{y}|\theta)$ is the likelihood of the data
$\mathbf{y}$ given $\theta$, and $f(\theta)$ is the prior for $\theta$.
In this case, the likelihood is the binomial likelihood that describes
the probability of a number of successes $y$ in $n$ Bernoulli trials
each with probability $\theta$ of success
\begin{eqnarray}
f(y|\theta) = \left(\begin{array}{cc} n \\ y \end{array} \right)
              \theta^{y} \left(1-\theta\right)^{n-y}.
\end{eqnarray}
As shown by \citet{sch14}, in this situation using a
Beta($\alpha$,$\beta$) prior on $\theta$ with hyperparameters $\alpha$
and $\beta$ results in a Beta posterior for $\theta$ of the form
Beta($\alpha+N_{\ast,\text{IRMP}},\beta+N_{\ast,\text{tot}}-N_{\ast,\text{IRMP}}$).

The hyperparameters $\alpha$ and $\beta$ of the prior can be thought
of as encoding a certain amount of prior information in the form
of pseudo-observations.  Specifically, $\alpha-1$ is the number of
success and $\beta-1$ is the number of failures imagined to have already
been observed and therefore included as prior information on $\theta$.
Taking any $\alpha = \beta = i$ where $i \geq 1$ could be thought of as
an uninformative prior in the sense that the probability of success and
failure in the prior distribution are equally likely.  However, if $i$
is large then there is imagined to be a lot of prior information and the
posterior distribution will mostly reflect the prior when $n \leq i$.
On the other hand, if $n \gg i$, then the posterior will be dominated
by the data.  For that reason, we take $\alpha=\beta=1$.

We provide the posterior median occurrence of IRMP stars in each
galactic environment in Table \ref{tab02}.  We define the lower
uncertainty as the difference between the posterior median and its
16th percentile.  Likewise, we define the upper uncertainty as the
difference between the posterior's 84th percentile and its median.
For an environment with no IRMP stars, we report 1-$\sigma$ upper
limits as the 68th percentile of the posterior distribution.
We find that $\eta_{\text{dSph}}=0.07_{-0.02}^{+0.02}$,
$\eta_{\text{Mag}}=0.037_{-0.006}^{+0.007}$,
$\eta_{\text{MWF}}=0.0013_{-0.0005}^{+0.0006}$, and a 1-$\sigma$ upper
limit $\eta_{\text{MWGC}}<0.00057$.  In words, IRMP stars are much more
common in the Milky Way's classical dSph satellites and the Magellanic
Clouds than in the Milky Way field or globular cluster populations.
IRMP stars are less common in globular clusters than in any other
galactic environment.  We find that the overlap probability between
the IRMP occurrences we observed in the Milky Way's classical dSph
satellites and the Magellanic Clouds is about one in 91, equivalent to
about 2.3 $\sigma$.  We find that the overlap probabilities between the
IRMP occurrences we observe in the Milky Way's classical dSph satellites
\& the Magellanic Clouds and the occurrence of IRMP stars we observe in
the Milky Way field population are about one in $2.8 \times 10^{15}$ and
one in $6.2 \times 10^{19}$, equivalent to about 8.1 and 9.2 $\sigma$.
The overlap probability between the IRMP occurrences we observe in the
Milky Way field and globular cluster populations is about one in 23,
equivalent to about 1.7 $\sigma$.  We summarize these occurrence posterior
overlap probabilities in Table \ref{tab03}.

\begin{deluxetable*}{ccC}
\tablecaption{Occurrence Posterior Overlap Probabilities\label{tab03}}
\tablewidth{0pt}
\tablehead{
\colhead{Population} &
\colhead{Population} &
\colhead{$P(\text{Posterior Overlap})$}
}
\startdata
dSph galaxies & Magellanic Clouds & 1.095 \times 10^{-2} \\
dSph galaxies & Milky Way field & 3.596 \times 10^{-16} \\
Magellanic Clouds & Milky Way field & 1.620 \times 10^{-20} \\
Milky Way field & Milky Way globular clusters & 4.300 \times 10^{-2}
\enddata
\end{deluxetable*}

\begin{figure*}
\plotone{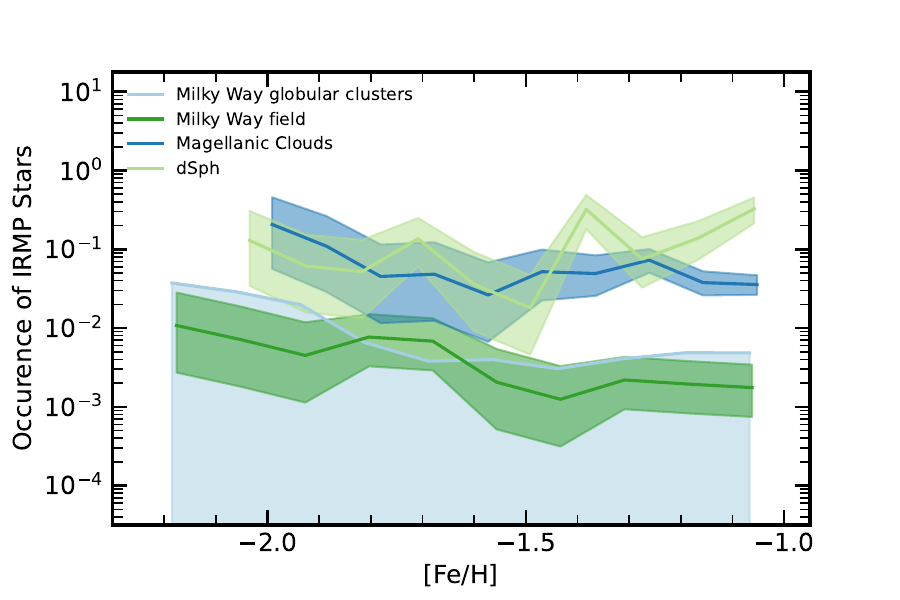}
\caption{Occurrence of iron-rich metal-poor stars as a function of
metallicity in different environments.  For the Milky Way field, dSph,
and Magellanic Cloud samples we plot the posterior medians of occurrence
as solid lines and the ranges defined by the 16th and 84th percentiles
of the occurrence posteriors as transparent polygons.  For the globular
cluster sample for which we infer an upper limit on IRMP occurrence, we
plot the 68th percentile of the posterior as a solid line and the range
defined by the 0th and 68th percentiles of the occurrence posterior as a
transparent polygon. There is no obvious dependence of IRMP occurrence on
metallicity in the Milky Way's field population, its globular clusters,
its classical dSph galaxies, or the Magellanic Clouds.\label{fig01}}
\end{figure*}

The results presented in Table \ref{tab02} are averaged over metallicity.
To investigate IRMP occurrence as a function of metallicity, for each
class of galactic environment we divide into ten equal intervals the
metallicity range spanned by our analysis sample for that environment.
We then apply the same occurrence formalism in each individual metallicity
interval and plot occurrence as a function of metallicity for each
environment in Figure \ref{fig01}.  We find no significant dependence
of the occurrence of IRMP stars on metallicity in our analysis sample.

\section{Discussion}\label{discussion}

We find that the occurrences of IRMP stars in the Milky Way's satellite
classical dSph galaxies, the Magellanic Clouds, the Milky Way field
population, and the Milky Way's globular cluster populations
have the values $\eta_{\text{dSph}}=0.07_{-0.02}^{+0.02}$,
$\eta_{\text{Mag}}=0.037_{-0.006}^{+0.007}$,
$\eta_{\text{MWF}}=0.0013_{-0.0005}^{+0.0006}$, and
$\eta_{\text{MWGC}}<0.00057$.  The probability that the IRMP occurrences
in the Milky Way's satellite classical dSph galaxies and the Magellanic
Clouds overlap is about one in 91, equivalent to about 2.3 $\sigma$.
The probabilities that the IRMP occurrence posterior for the Milky
Way field overlaps with the IRMP occurrence posteriors for Milky Way's
satellite classical dSph galaxies and the Magellanic Clouds are about
one in $2.8 \times 10^{15}$ and one in $6.2 \times 10^{19}$, equivalent
to about 8.1 and 9.2 $\sigma$.  The probability that the IRMP occurrences
in the Milky Way field and globular cluster populations overlap is about
one in 23, equivalent to about 1.7 $\sigma$.  While the absolute values of
IRMP star occurrences may be difficult to interpret, as we argued in the
introduction their ordering $\eta_{\text{MWGC}} \sim \eta_{\text{MWF}} <
\eta_{\text{Mag}} \sim \eta_{\text{dSph}}$ has two important implications.

The increased occurrence of IRMP stars in environments like the Milky
Way's satellite classical dSph galaxies and the Magellanic Clouds where
thermonuclear supernovae were important contributors to chemical evolution
supports the \citet{reg23} scenario for IRMP star formation in molecular
cores with metal contents dominated by the thermonuclear supernova
nucleosynthesis.  The confirmation of this \citet{reg23} prediction
reinforces the idea that the elemental abundances of individual IRMP
stars can be used to investigate the progenitor systems and explosion
mechanisms responsible for the thermonuclear supernovae that produced
much of their metal contents.

The statistically indistinguishable occurrences of IRMP stars in the Milky
Way's field and globular cluster populations suggests that the time lags
between the formation of globular clusters' first and second stellar
generations were longer than the thermonuclear supernova delay time.
This is broadly consistent with the idea from the asymptotic giant
branch (AGB) scenario for globular cluster multiple populations that
the explosions of thermonuclear supernovae associated with a globular
cluster's first stellar generation quench the star formation event that
produced its second stellar generation \citep[e.g.,][]{der08,cal19,lal21}.
Given the suspected importance of thermonuclear supernovae for quenching
second generation star formation in globular clusters, it remains to
be explained why globular cluster second generation stars that bear the
imprint of thermonuclear supernova nucleosynthesis are so rare as to not
appear in a sample of nearly 2000 globular cluster members.  Our result is
also consistent with scenarios for globular cluster multiple populations
invoking a thermonuclear supernova at the start of a cluster's evolution
\citep[e.g.,][]{mar09,san12}.

Contributions from both core-collapse and thermonuclear supernovae as well
as $s$- and $r$-process nucleosynthesis are required to explain the solar
abundance pattern.  While the progenitors of core-collapse supernovae are
known to be massive stars and the $s$-process takes place in AGB stars,
the progenitors of thermonuclear supernovae and the astrophysical site of
the $r$-process are more uncertain.  Observational constraints on either
the progenitors of thermonuclear supernovae or the astrophysical site of
the $r$-process are therefore valuable.  The kilonova GW170817 confirmed
that neutron stars mergers are at least partially responsible for
$r$-process nucleosynthesis \citep[e.g.,][]{abb17,arc17}.  The relative
occurrences of IRMP and $r$-process enhanced stars in the same populations
can be used to compare the relative probabilities of the circumstances
that lead to the formation of IRMP and $r$-process enhanced stars.

We find that the occurrences of IRMP stars are an order of magnitude
lower than the occurrence of $r$-process enhanced stars in both the
Milky Way field and Magellanic Clouds.  In the Milky Way field,
we find that $\eta_{\text{IRMP,MWF}}=0.0013_{-0.0005}^{+0.0006}$
while \citet{bar05} found $\eta_{\text{rII,MWF}}\approx0.03$
according to the definition of highly $r$-process enhanced
(i.e., r-II stars) from \citet{bee05}.\footnote{More recent
estimates by the $r$-process Alliance have found similar occurrences
\citep[e.g.,][]{han18,sak18,ezz20}.} In the Magellanic Clouds, we find
that $\eta_{\text{IRMP,Mag}}=0.037_{-0.006}^{+0.007}$ while \citet{reg21}
found $\eta_{\text{rII,Mag}}=0.38^{+0.14}_{-0.13}$.  We conclude that
the circumstances that lead to the formation of IRMP stars occur an
order of magnitude less frequently than the circumstances that lead to
the formation of $r$-process enhanced stars.

\section{Conclusion}\label{conclusion}

We conclude that iron-rich metal-poor stars with $[\text{Fe/H}]
\lesssim -1$ and $[\text{O,Na,Mg,Al,K,Co/Fe}] < 0$ are more common in
the Milky Way's satellite classical dSph galaxies and the Magellanic
Clouds than in the Milky Way field or globular cluster populations.
Because thermonuclear supernovae are thought to have been more important
contributors to the chemical evolution of the Milky Way's satellite
classical dSph galaxies and the Magellanic Clouds than to the Milky
Way field and globular cluster populations in the range $[\text{Fe/H}]
\lesssim -1$, our inferences confirm the interpretation of iron-rich
metal-poor stars put forward in \citet{reg23} proposing that iron-rich
metal-poor stars formed from molecular cores with metal contents dominated
by thermonuclear supernova nucleosynthesis.  Iron-rich metal-poor stars
can therefore be used to constrain the progenitor systems and explosion
mechanisms of the thermonuclear supernovae responsible for their elemental
abundances.  We further find that the occurrences of iron-rich metal-poor
stars in the Milky Way's field and globular cluster populations are
statistically indistinguishable.  This observation implies that the time
lag between the formation of a globular cluster's first and second stellar
generations was longer than the thermonuclear supernova delay time.
It is also consistent with explanations for globular cluster multiple
populations that require early enrichment by thermonuclear supernovae.

\section*{Acknowledgments}

We thank Yossef Zenati for sharing his expertise on Type Ia supernovae.
Support for this work was provided by Johns Hopkins University
through a Summer Provost's Undergraduate Research Award to Zachary
Reeves and by the Carnegie Institution for Science through a Carnegie
Postdoctoral Fellowship award to Henrique Reggiani.  Funding for
SDSS-III has been provided by the Alfred P. Sloan Foundation,
the Participating Institutions, the National Science Foundation,
and the U.S. Department of Energy Office of Science. The SDSS-III
web site is \url{http://www.sdss3.org/}.  SDSS-III is managed by the
Astrophysical Research Consortium for the Participating Institutions
of the SDSS-III Collaboration including the University of Arizona, the
Brazilian Participation Group, Brookhaven National Laboratory, Carnegie
Mellon University, University of Florida, the French Participation
Group, the German Participation Group, Harvard University, the
Instituto de Astrofisica de Canarias, the Michigan State/Notre
Dame/JINA Participation Group, Johns Hopkins University, Lawrence
Berkeley National Laboratory, Max Planck Institute for Astrophysics,
Max Planck Institute for Extraterrestrial Physics, New Mexico State
University, New York University, Ohio State University, Pennsylvania
State University, University of Portsmouth, Princeton University, the
Spanish Participation Group, University of Tokyo, University of Utah,
Vanderbilt University, University of Virginia, University of Washington,
and Yale University.  Funding for the Sloan Digital Sky Survey IV has
been provided by the Alfred P. Sloan Foundation, the U.S.  Department
of Energy Office of Science, and the Participating Institutions.
SDSS-IV acknowledges support and resources from the Center for High
Performance Computing at the University of Utah. The SDSS website is
\url{www.sdss.org}.  SDSS-IV is managed by the Astrophysical Research
Consortium for the Participating Institutions of the SDSS Collaboration
including the Brazilian Participation Group, the Carnegie Institution
for Science, Carnegie Mellon University, Center for Astrophysics |
Harvard \& Smithsonian, the Chilean Participation Group, the French
Participation Group, Instituto de Astrof\'isica de Canarias, The Johns
Hopkins University, Kavli Institute for the Physics and Mathematics of
the Universe (IPMU) / University of Tokyo, the Korean Participation
Group, Lawrence Berkeley National Laboratory, Leibniz Institut f\"ur
Astrophysik Potsdam (AIP),  Max-Planck-Institut f\"ur Astronomie (MPIA
Heidelberg), Max-Planck-Institut f\"ur Astrophysik (MPA Garching),
Max-Planck-Institut f\"ur Extraterrestrische Physik (MPE), National
Astronomical Observatories of China, New Mexico State University, New
York University, University of Notre Dame, Observat\'ario Nacional /
MCTI, The Ohio State University, Pennsylvania State University, Shanghai
Astronomical Observatory, United Kingdom Participation Group, Universidad
Nacional Aut\'onoma de M\'exico, University of Arizona, University
of Colorado Boulder, University of Oxford, University of Portsmouth,
University of Utah, University of Virginia, University of Washington,
University of Wisconsin, Vanderbilt University, and Yale University.
This work has made use of data from the European Space Agency (ESA)
mission {\it Gaia} (\url{https://www.cosmos.esa.int/gaia}), processed
by the {\it Gaia} Data Processing and Analysis Consortium (DPAC,
\url{https://www.cosmos.esa.int/web/gaia/dpac/consortium}).  Funding for
the DPAC has been provided by national institutions, in particular the
institutions participating in the {\it Gaia} Multilateral Agreement.
This publication makes use of data products from the Two Micron All
Sky Survey, which is a joint project of the University of Massachusetts
and the Infrared Processing and Analysis Center/California Institute of
Technology, funded by the National Aeronautics and Space Administration
and the National Science Foundation.  This research has made use of the
SIMBAD database, operated at CDS, Strasbourg, France \citep{wen00}.
This research has made use of the VizieR catalog access tool, CDS,
Strasbourg, France.  The original description of the VizieR service
was published in \citet{och00}.  This research has made use of NASA's
Astrophysics Data System Bibliographic Services.


\vspace{5mm}
\facilities{CDS, CTIO:2MASS, Du Pont (APOGEE), FLWO:2MASS, Gaia, Sloan
(APOGEE)}

\software{\texttt{astropy} \citep{astropy13,astropy18},
          \texttt{matplotlib} \citep{hun07}
          \texttt{numpy} \citep{har20},
          \texttt{pandas} \citep{mck10,reb20},
          \texttt{R} \citep{r23},
          \texttt{scipy} \citep{vir20}
          }

\appendix
SDSS DR17 APOGEE data quality and targeting information are stored as
bitmasks.\footnote{\url{https://www.sdss.org/dr17/irspec/apogee-bitmasks}}
Since unreliable photospheric stellar parameters will lead to unreliable
elemental abundances, we exclude from our analysis stars with the bits
\texttt{TEFF\_WARN}, \texttt{LOGG\_WARN}, \texttt{VMICRO\_WARN},
\texttt{M\_H\_WARN}, \texttt{STAR\_WARN}, \texttt{CHI2\_WARN},
\texttt{COLORTE\_WARN}, \texttt{ROTATION\_WARN}, \texttt{SN\_WARN},
\texttt{TEFF\_BAD}, \texttt{LOGG\_BAD}, \texttt{VMICRO\_BAD},
\texttt{M\_H\_BAD}, \texttt{STAR\_BAD}, \texttt{CHI2\_BAD},
\texttt{COLORTE\_BAD}, \texttt{ROTATION\_BAD}, or \texttt{SN\_BAD} set
in the column \texttt{aspcapflag} in the table \texttt{aspcapStar}.
These flags correspond to binary digits 0, 1, 2, 3, 7, 8, 9, 10,
11, 16, 17, 18, 19, 23, 24, 25, 26, and 27.  Note that though the
bit corresponding to \texttt{STAR\_WARN} should be set if any of the
bits \texttt{TEFF\_WARN}, \texttt{LOGG\_WARN}, \texttt{CHI2\_WARN},
\texttt{COLORTE\_WARN}, \texttt{ROTATION\_WARN}, or \texttt{SN\_WARN} are
set we choose to include all of these bits in our data quality checks.
Likewise, though the bit corresponding to \texttt{STAR\_BAD} should
be set if any of the bits \texttt{TEFF\_BAD}, \texttt{LOGG\_BAD},
\texttt{CHI2\_BAD}, \texttt{COLORTE\_BAD}, \texttt{ROTATION\_BAD},
or \texttt{SN\_BAD} are set we choose to include all of these bits in
our data quality checks.  We remove duplicate observations from our
analysis sample by rejecting objects with binary digit 4 set in the
column \texttt{extratarg} in the table \texttt{apogeeStar}.

We exclude from our analysis any elemental abundances that
are indicated as suspect.  We set to \texttt{null} all
elemental abundances with the final value $-9999$ or with
any of the bits \texttt{GRIDEDGE\_BAD}, \texttt{CALRANGE\_BAD},
\texttt{OTHER\_BAD}, \texttt{FERRE\_FAIL}, \texttt{PARAM\_MISMATCH\_BAD},
\texttt{TEFF\_CUT}, \texttt{GRIDEDGE\_WARN}, \texttt{CALRANGE\_WARN},
\texttt{OTHER\_WARN}, \texttt{PARAM\_MISMATCH\_WARN}, \texttt{ERR\_WARN},
or \texttt{PARAM\_FIXED} set in a column \texttt{*\_fe\_flag} in the
table \texttt{aspcapStar}.  These flags correspond to binary digits 0,
1, 2, 3, 4, 6, 8, 9, 10, 12, 14, and 16.  We ultimately use the following
query in the CasJobs portal to generate our analysis sample.
\\
\begin{verbatim}
SELECT a.apogee_id, b.ra, b.dec, b.glon, b.glat, b.snr, b.extratarg,
CASE WHEN (((a.teff_flag & 87903) = 0) AND (a.teff > -9999))
   THEN a.teff ELSE null END AS teff,
CASE WHEN (((a.teff_flag & 87903) = 0) AND (a.teff > -9999))
   THEN a.teff_err ELSE null END AS teff_err,
CASE WHEN (((a.logg_flag & 87903) = 0) AND (a.logg > -9999))
   THEN a.logg ELSE null END AS logg,
CASE WHEN (((a.logg_flag & 87903) = 0) AND (a.logg > -9999))
   THEN a.logg_err ELSE null END AS logg_err,
CASE WHEN (((a.m_h_flag & 87903) = 0) AND (a.m_h > -9999))
   THEN a.m_h ELSE null END AS m_h,
CASE WHEN (((a.m_h_flag & 87903) = 0) AND (a.m_h > -9999))
   THEN a.m_h_err ELSE null END AS m_h_err,
CASE WHEN (((a.fe_h_flag & 87903) = 0) AND (fe_h > -9999))
   THEN fe_h ELSE null END AS fe_h,
CASE WHEN (((a.fe_h_flag & 87903) = 0) AND (fe_h > -9999))
   THEN fe_h_err ELSE null END AS fe_h_err,
CASE WHEN (((a.o_fe_flag & 87903) = 0) AND ((a.fe_h_flag & 87903) = 0)
   AND (a.o_fe > -9999) AND (a.fe_h > -9999))
   THEN a.o_fe ELSE null END AS o_fe,
CASE WHEN (((a.o_fe_flag & 87903) = 0) AND ((a.fe_h_flag & 87903) = 0)
   AND (a.o_fe > -9999) AND (a.fe_h > -9999))
   THEN a.o_fe_err ELSE null END AS o_fe_err,
CASE WHEN (((a.na_fe_flag & 87903) = 0) AND ((a.fe_h_flag & 87903) = 0)
   AND (a.na_fe > -9999) AND (a.fe_h > -9999))
   THEN a.na_fe ELSE null END AS na_fe,
CASE WHEN (((a.na_fe_flag & 87903) = 0) AND ((a.fe_h_flag & 87903) = 0)
   AND (a.na_fe > -9999) AND (a.fe_h > -9999))
   THEN a.na_fe_err ELSE null END AS na_fe_err,
CASE WHEN (((a.mg_fe_flag & 87903) = 0) AND ((a.fe_h_flag & 87903) = 0)
   AND (a.mg_fe > -9999) AND (a.fe_h > -9999))
   THEN a.mg_fe ELSE null END AS mg_fe,
CASE WHEN (((a.mg_fe_flag & 87903) = 0) AND ((a.fe_h_flag & 87903) = 0)
   AND (a.mg_fe > -9999) AND (a.fe_h > -9999))
   THEN a.mg_fe_err ELSE null END AS mg_fe_err,
CASE WHEN (((a.al_fe_flag & 87903) = 0) AND ((a.fe_h_flag & 87903) = 0)
   AND (a.al_fe > -9999) AND (a.fe_h > -9999))
   THEN a.al_fe ELSE null END AS al_fe,
CASE WHEN (((a.al_fe_flag & 87903) = 0) AND ((a.fe_h_flag & 87903) = 0)
   AND (a.al_fe > -9999) AND (a.fe_h > -9999))
   THEN a.al_fe_err ELSE null END AS al_fe_err,
CASE WHEN (((a.k_fe_flag & 87903) = 0) AND ((a.fe_h_flag & 87903) = 0)
   AND (a.k_fe > -9999) AND (a.fe_h > -9999))
   THEN a.k_fe ELSE null END AS k_fe,
CASE WHEN (((a.k_fe_flag & 87903) = 0) AND ((a.fe_h_flag & 87903) = 0)
   AND (a.k_fe > -9999) AND (a.fe_h > -9999))
   THEN a.k_fe_err ELSE null END AS k_fe_err,
CASE WHEN (((a.co_fe_flag & 87903) = 0) AND ((a.fe_h_flag & 87903) = 0)
   AND (a.co_fe > -9999) AND (a.fe_h > -9999))
   THEN a.co_fe ELSE null END AS co_fe,
CASE WHEN (((a.co_fe_flag & 87903) = 0) AND ((a.fe_h_flag & 87903) = 0)
   AND (a.co_fe > -9999) AND (a.fe_h > -9999))
   THEN a.co_fe_err ELSE null END AS co_fe_err
FROM DR17.aspcapStar a
INNER JOIN DR17.apogeeStar b ON a.apstar_id = b.apstar_id
WHERE (a.aspcapflag & 261033871) = 0
AND (b.extratarg & 16) = 0
AND a.logg < 3.8
\end{verbatim}

\bibliography{ms}{}

\begin{thebibliography}{}
\expandafter\ifx\csname natexlab\endcsname\relax\def\natexlab#1{#1}\fi
\providecommand{\url}[1]{\href{#1}{#1}}
\providecommand{\dodoi}[1]{doi:~\href{http://doi.org/#1}{\nolinkurl{#1}}}
\providecommand{\doeprint}[1]{\href{http://ascl.net/#1}{\nolinkurl{http://ascl.net/#1}}}
\providecommand{\doarXiv}[1]{\href{https://arxiv.org/abs/#1}{\nolinkurl{https://arxiv.org/abs/#1}}}

\bibitem[{{Abbott} {et~al.}(2017){Abbott}, {Abbott}, {Abbott}, {Acernese},
  {Ackley}, {Adams}, {Adams}, {Addesso}, {Adhikari}, {Adya}, {Affeldt},
  {Afrough}, {Agarwal}, {Agathos}, {Agatsuma}, {Aggarwal}, {Aguiar}, {Aiello},
  {Ain}, {Ajith}, {Allen}, {Allen}, {Allocca}, {Altin}, {Amato}, {Ananyeva},
  {Anderson}, {Anderson}, {Angelova}, {Antier}, {Appert}, {Arai}, {Araya},
  {Areeda}, {Arnaud}, {Arun}, {Ascenzi}, {Ashton}, {Ast}, {Aston}, {Astone},
  {Atallah}, {Aufmuth}, {Aulbert}, {AultONeal}, {Austin}, {Avila-Alvarez},
  {Babak}, {Bacon}, {Bader}, {Bae}, {Bailes}, {Baker}, {Baldaccini},
  {Ballardin}, {Ballmer}, {Banagiri}, {Barayoga}, {Barclay}, {Barish},
  {Barker}, {Barkett}, {Barone}, {Barr}, {Barsotti}, {Barsuglia}, {Barta},
  {Barthelmy}, {Bartlett}, {Bartos}, {Bassiri}, {Basti}, {Batch}, {Bawaj},
  {Bayley}, {Bazzan}, {B{\'e}csy}, {Beer}, {Bejger}, {Belahcene}, {Bell},
  {Berger}, {Bergmann}, {Bernuzzi}, {Bero}, {Berry}, {Bersanetti}, {Bertolini},
  {Betzwieser}, {Bhagwat}, {Bhandare}, {Bilenko}, {Billingsley}, {Billman},
  {Birch}, {Birney}, {Birnholtz}, {Biscans}, {Biscoveanu}, {Bisht}, {Bitossi},
  {Biwer}, {Bizouard}, {Blackburn}, {Blackman}, {Blair}, {Blair}, {Blair},
  {Bloemen}, {Bock}, {Bode}, {Boer}, {Bogaert}, {Bohe}, {Bondu}, {Bonilla},
  {Bonnand}, {Boom}, {Bork}, {Boschi}, {Bose}, {Bossie}, {Bouffanais}, {Bozzi},
  {Bradaschia}, {Brady}, {Branchesi}, {Brau}, {Briant}, {Brillet}, {Brinkmann},
  {Brisson}, {Brockill}, {Broida}, {Brooks}, {Brown}, {Brown}, {Brunett},
  {Buchanan}, {Buikema}, {Bulik}, {Bulten}, {Buonanno}, {Buskulic}, {Buy},
  {Byer}, {Cabero}, {Cadonati}, {Cagnoli}, {Cahillane}, {Calder{\'o}n
  Bustillo}, {Callister}, {Calloni}, {Camp}, {Canepa}, {Canizares}, {Cannon},
  {Cao}, {Cao}, {Capano}, {Capocasa}, {Carbognani}, {Caride}, {Carney},
  {Carullo}, {Casanueva Diaz}, {Casentini}, {Caudill}, {Cavagli{\`a}},
  {Cavalier}, {Cavalieri}, {Cella}, {Cepeda}, {Cerd{\'a}-Dur{\'a}n},
  {Cerretani}, {Cesarini}, {Chamberlin}, {Chan}, {Chao}, {Charlton}, {Chase},
  {Chassande-Mottin}, {Chatterjee}, {Chatziioannou}, {Cheeseboro}, {Chen},
  {Chen}, {Chen}, {Cheng}, {Chia}, {Chincarini}, {Chiummo}, {Chmiel}, {Cho},
  {Cho}, {Chow}, {Christensen}, {Chu}, {Chua}, {Chua}, {Chung}, {Chung},
  {Ciani}, {Ciolfi}, {Cirelli}, {Cirone}, {Clara}, {Clark}, {Clearwater},
  {Cleva}, {Cocchieri}, {Coccia}, {Cohadon}, {Cohen}, {Colla}, {Collette},
  {Cominsky}, {Constancio}, {Conti}, {Cooper}, {Corban}, {Corbitt},
  {Cordero-Carri{\'o}n}, {Corley}, {Cornish}, {Corsi}, {Cortese}, {Costa},
  {Coughlin}, {Coughlin}, {Coulon}, {Countryman}, {Couvares}, {Covas}, {Cowan},
  {Coward}, {Cowart}, {Coyne}, {Coyne}, {Creighton}, {Creighton}, {Cripe},
  {Crowder}, {Cullen}, {Cumming}, {Cunningham}, {Cuoco}, {Dal Canton},
  {D{\'a}lya}, {Danilishin}, {D'Antonio}, {Danzmann}, {Dasgupta}, {Da Silva
  Costa}, {Dattilo}, {Dave}, {Davier}, {Davis}, {Daw}, {Day}, {De}, {DeBra},
  {Degallaix}, {De Laurentis}, {Del{\'e}glise}, {Del Pozzo}, {Demos}, {Denker},
  {Dent}, {De Pietri}, {Dergachev}, {De Rosa}, {DeRosa}, {De Rossi}, {DeSalvo},
  {de Varona}, {Devenson}, {Dhurandhar}, {D{\'\i}az}, {Dietrich}, {Di Fiore},
  {Di Giovanni}, {Di Girolamo}, {Di Lieto}, {Di Pace}, {Di Palma}, {Di Renzo},
  {Doctor}, {Dolique}, {Donovan}, {Dooley}, {Doravari}, {Dorrington},
  {Douglas}, {Dovale {\'A}lvarez}, {Downes}, {Drago}, {Dreissigacker},
  {Driggers}, {Du}, {Ducrot}, {Dudi}, {Dupej}, {Dwyer}, {Edo}, {Edwards},
  {Effler}, {Eggenstein}, {Ehrens}, {Eichholz}, {Eikenberry}, {Eisenstein},
  {Essick}, {Estevez}, {Etienne}, {Etzel}, {Evans}, {Evans}, {Factourovich},
  {Fafone}, {Fair}, {Fairhurst}, {Fan}, {Farinon}, {Farr}, {Farr},
  {Fauchon-Jones}, {Favata}, {Fays}, {Fee}, {Fehrmann}, {Feicht}, {Fejer},
  {Fernandez-Galiana}, {Ferrante}, {Ferreira}, {Ferrini}, {Fidecaro},
  {Finstad}, {Fiori}, {Fiorucci}, {Fishbach}, {Fisher}, {Fitz-Axen},
  {Flaminio}, {Fletcher}, {Fong}, {Font}, {Forsyth}, {Forsyth}, {Fournier},
  {Frasca}, {Frasconi}, {Frei}, {Freise}, {Frey}, {Frey}, {Fries}, {Fritschel},
  {Frolov}, {Fulda}, {Fyffe}, {Gabbard}, {Gadre}, {Gaebel}, {Gair},
  {Gammaitoni}, {Ganija}, {Gaonkar}, {Garcia-Quiros}, {Garufi}, {Gateley},
  {Gaudio}, {Gaur}, {Gayathri}, {Gehrels}, {Gemme}, {Genin}, {Gennai},
  {George}, {George}, {Gergely}, {Germain}, {Ghonge}, {Ghosh}, {Ghosh},
  {Ghosh}, {Giaime}, {Giardina}, {Giazotto}, {Gill}, {Glover}, {Goetz},
  {Goetz}, {Gomes}, {Goncharov}, {Gonz{\'a}lez}, {Gonzalez Castro},
  {Gopakumar}, {Gorodetsky}, {Gossan}, {Gosselin}, {Gouaty}, {Grado}, {Graef},
  {Granata}, {Grant}, {Gras}, {Gray}, {Greco}, {Green}, {Gretarsson}, {Groot},
  {Grote}, {Grunewald}, {Gruning}, {Guidi}, {Guo}, {Gupta}, {Gupta}, {Gushwa},
  {Gustafson}, {Gustafson}, {Halim}, {Hall}, {Hall}, {Hamilton}, {Hammond},
  {Haney}, {Hanke}, {Hanks}, {Hanna}, {Hannam}, {Hannuksela}, {Hanson},
  {Hardwick}, {Harms}, {Harry}, {Harry}, {Hart}, {Haster}, {Haughian}, {Healy},
  {Heidmann}, {Heintze}, {Heitmann}, {Hello}, {Hemming}, {Hendry}, {Heng},
  {Hennig}, {Heptonstall}, {Heurs}, {Hild}, {Hinderer}, {Ho}, {Hoak}, {Hofman},
  {Holt}, {Holz}, {Hopkins}, {Horst}, {Hough}, {Houston}, {Howell}, {Hreibi},
  {Hu}, {Huerta}, {Huet}, {Hughey}, {Husa}, {Huttner}, {Huynh-Dinh}, {Indik},
  {Inta}, {Intini}, {Isa}, {Isac}, {Isi}, {Iyer}, {Izumi}, {Jacqmin}, {Jani},
  {Jaranowski}, {Jawahar}, {Jim{\'e}nez-Forteza}, {Johnson},
  {Johnson-McDaniel}, {Jones}, {Jones}, {Jonker}, {Ju}, {Junker}, {Kalaghatgi},
  {Kalogera}, {Kamai}, {Kandhasamy}, {Kang}, {Kanner}, {Kapadia}, {Karki},
  {Karvinen}, {Kasprzack}, {Kastaun}, {Katolik}, {Katsavounidis}, {Katzman},
  {Kaufer}, {Kawabe}, {K{\'e}f{\'e}lian}, {Keitel}, {Kemball}, {Kennedy},
  {Kent}, {Key}, {Khalili}, {Khan}, {Khan}, {Khan}, {Khazanov}, {Kijbunchoo},
  {Kim}, {Kim}, {Kim}, {Kim}, {Kim}, {Kim}, {Kimbrell}, {King}, {King},
  {Kinley-Hanlon}, {Kirchhoff}, {Kissel}, {Kleybolte}, {Klimenko}, {Knowles},
  {Koch}, {Koehlenbeck}, {Koley}, {Kondrashov}, {Kontos}, {Korobko}, {Korth},
  {Kowalska}, {Kozak}, {Kr{\"a}mer}, {Kringel}, {Krishnan}, {Kr{\'o}lak},
  {Kuehn}, {Kumar}, {Kumar}, {Kumar}, {Kuo}, {Kutynia}, {Kwang}, {Lackey},
  {Lai}, {Landry}, {Lang}, {Lange}, {Lantz}, {Lanza}, {Larson},
  {Lartaux-Vollard}, {Lasky}, {Laxen}, {Lazzarini}, {Lazzaro}, {Leaci},
  {Leavey}, {Lee}, {Lee}, {Lee}, {Lee}, {Lee}, {Lehmann}, {Lenon}, {Leon},
  {Leonardi}, {Leroy}, {Letendre}, {Levin}, {Li}, {Linker}, {Littenberg},
  {Liu}, {Liu}, {Lo}, {Lockerbie}, {London}, {Lord}, {Lorenzini}, {Loriette},
  {Lormand}, {Losurdo}, {Lough}, {Lousto}, {Lovelace}, {L{\"u}ck}, {Lumaca},
  {Lundgren}, {Lynch}, {Ma}, {Macas}, {Macfoy}, {Machenschalk}, {MacInnis},
  {Macleod}, {Maga{\~n}a Hernandez}, {Maga{\~n}a-Sandoval}, {Maga{\~n}a
  Zertuche}, {Magee}, {Majorana}, {Maksimovic}, {Man}, {Mandic}, {Mangano},
  {Mansell}, {Manske}, {Mantovani}, {Marchesoni}, {Marion}, {M{\'a}rka},
  {M{\'a}rka}, {Markakis}, {Markosyan}, {Markowitz}, {Maros}, {Marquina},
  {Marsh}, {Martelli}, {Martellini}, {Martin}, {Martin}, {Martynov}, {Marx},
  {Mason}, {Massera}, {Masserot}, {Massinger}, {Masso-Reid}, {Mastrogiovanni},
  {Matas}, {Matichard}, {Matone}, {Mavalvala}, {Mazumder}, {McCarthy},
  {McClelland}, {McCormick}, {McCuller}, {McGuire}, {McIntyre}, {McIver},
  {McManus}, {McNeill}, {McRae}, {McWilliams}, {Meacher}, {Meadors}, {Mehmet},
  {Meidam}, {Mejuto-Villa}, {Melatos}, {Mendell}, {Mercer}, {Merilh},
  {Merzougui}, {Meshkov}, {Messenger}, {Messick}, {Metzdorff}, {Meyers},
  {Miao}, {Michel}, {Middleton}, {Mikhailov}, {Milano}, {Miller}, {Miller},
  {Miller}, {Millhouse}, {Milovich-Goff}, {Minazzoli}, {Minenkov}, {Ming},
  {Mishra}, {Mitra}, {Mitrofanov}, {Mitselmakher}, {Mittleman}, {Moffa},
  {Moggi}, {Mogushi}, {Mohan}, {Mohapatra}, {Molina}, {Montani}, {Moore},
  {Moraru}, {Moreno}, {Morisaki}, {Morriss}, {Mours}, {Mow-Lowry}, {Mueller},
  {Muir}, {Mukherjee}, {Mukherjee}, {Mukherjee}, {Mukund}, {Mullavey}, {Munch},
  {Mu{\~n}iz}, {Muratore}, {Murray}, {Nagar}, {Napier}, {Nardecchia},
  {Naticchioni}, {Nayak}, {Neilson}, {Nelemans}, {Nelson}, {Nery}, {Neunzert},
  {Nevin}, {Newport}, {Newton}, {Ng}, {Nguyen}, {Nguyen}, {Nichols}, {Nielsen},
  {Nissanke}, {Nitz}, {Noack}, {Nocera}, {Nolting}, {North}, {Nuttall},
  {Oberling}, {O'Dea}, {Ogin}, {Oh}, {Oh}, {Ohme}, {Okada}, {Oliver},
  {Oppermann}, {Oram}, {O'Reilly}, {Ormiston}, {Ortega}, {O'Shaughnessy},
  {Ossokine}, {Ottaway}, {Overmier}, {Owen}, {Pace}, {Page}, {Page}, {Pai},
  {Pai}, {Palamos}, {Palashov}, {Palomba}, {Pal-Singh}, {Pan}, {Pan}, {Pang},
  {Pang}, {Pankow}, {Pannarale}, {Pant}, {Paoletti}, {Paoli}, {Papa}, {Parida},
  {Parker}, {Pascucci}, {Pasqualetti}, {Passaquieti}, {Passuello}, {Patil},
  {Patricelli}, {Pearlstone}, {Pedraza}, {Pedurand}, {Pekowsky}, {Pele},
  {Penn}, {Perez}, {Perreca}, {Perri}, {Pfeiffer}, {Phelps}, {Piccinni},
  {Pichot}, {Piergiovanni}, {Pierro}, {Pillant}, {Pinard}, {Pinto}, {Pirello},
  {Pitkin}, {Poe}, {Poggiani}, {Popolizio}, {Porter}, {Post}, {Powell},
  {Prasad}, {Pratt}, {Pratten}, {Predoi}, {Prestegard}, {Prijatelj},
  {Principe}, {Privitera}, {Prix}, {Prodi}, {Prokhorov}, {Puncken}, {Punturo},
  {Puppo}, {P{\"u}rrer}, {Qi}, {Quetschke}, {Quintero}, {Quitzow-James},
  {Raab}, {Rabeling}, {Radkins}, {Raffai}, {Raja}, {Rajan}, {Rajbhandari},
  {Rakhmanov}, {Ramirez}, {Ramos-Buades}, {Rapagnani}, {Raymond}, {Razzano},
  {Read}, {Regimbau}, {Rei}, {Reid}, {Reitze}, {Ren}, {Reyes}, {Ricci},
  {Ricker}, {Rieger}, {Riles}, {Rizzo}, {Robertson}, {Robie}, {Robinet},
  {Rocchi}, {Rolland}, {Rollins}, {Roma}, {Romano}, {Romano}, {Romel}, {Romie},
  {Rosi{\'n}ska}, {Ross}, {Rowan}, {R{\"u}diger}, {Ruggi}, {Rutins}, {Ryan},
  {Sachdev}, {Sadecki}, {Sadeghian}, {Sakellariadou}, {Salconi}, {Saleem},
  {Salemi}, {Samajdar}, {Sammut}, {Sampson}, {Sanchez}, {Sanchez},
  {Sanchis-Gual}, {Sandberg}, {Sanders}, {Sassolas}, {Sathyaprakash},
  {Saulson}, {Sauter}, {Savage}, {Sawadsky}, {Schale}, {Scheel}, {Scheuer},
  {Schmidt}, {Schmidt}, {Schnabel}, {Schofield}, {Sch{\"o}nbeck}, {Schreiber},
  {Schuette}, {Schulte}, {Schutz}, {Schwalbe}, {Scott}, {Scott}, {Seidel},
  {Sellers}, {Sengupta}, {Sentenac}, {Sequino}, {Sergeev}, {Shaddock},
  {Shaffer}, {Shah}, {Shahriar}, {Shaner}, {Shao}, {Shapiro}, {Shawhan},
  {Sheperd}, {Shoemaker}, {Shoemaker}, {Siellez}, {Siemens}, {Sieniawska},
  {Sigg}, {Silva}, {Singer}, {Singh}, {Singhal}, {Sintes}, {Slagmolen},
  {Smith}, {Smith}, {Smith}, {Somala}, {Son}, {Sonnenberg}, {Sorazu},
  {Sorrentino}, {Souradeep}, {Spencer}, {Srivastava}, {Staats}, {Staley},
  {Steinke}, {Steinlechner}, {Steinlechner}, {Steinmeyer}, {Stevenson},
  {Stone}, {Stops}, {Strain}, {Stratta}, {Strigin}, {Strunk}, {Sturani},
  {Stuver}, {Summerscales}, {Sun}, {Sunil}, {Suresh}, {Sutton}, {Swinkels},
  {Szczepa{\'n}czyk}, {Tacca}, {Tait}, {Talbot}, {Talukder}, {Tanner},
  {T{\'a}pai}, {Taracchini}, {Tasson}, {Taylor}, {Taylor}, {Tewari}, {Theeg},
  {Thies}, {Thomas}, {Thomas}, {Thomas}, {Thorne}, {Thorne}, {Thrane},
  {Tiwari}, {Tiwari}, {Tokmakov}, {Toland}, {Tonelli}, {Tornasi},
  {Torres-Forn{\'e}}, {Torrie}, {T{\"o}yr{\"a}}, {Travasso}, {Traylor},
  {Trinastic}, {Tringali}, {Trozzo}, {Tsang}, {Tse}, {Tso}, {Tsukada}, {Tsuna},
  {Tuyenbayev}, {Ueno}, {Ugolini}, {Unnikrishnan}, {Urban}, {Usman},
  {Vahlbruch}, {Vajente}, {Valdes}, {Vallisneri}, {van Bakel}, {van Beuzekom},
  {van den Brand}, {Van Den Broeck}, {Vander-Hyde}, {van der Schaaf}, {van
  Heijningen}, {van Veggel}, {Vardaro}, {Varma}, {Vass}, {Vas{\'u}th},
  {Vecchio}, {Vedovato}, {Veitch}, {Veitch}, {Venkateswara}, {Venugopalan},
  {Verkindt}, {Vetrano}, {Vicer{\'e}}, {Viets}, {Vinciguerra}, {Vine}, {Vinet},
  {Vitale}, {Vo}, {Vocca}, {Vorvick}, {Vyatchanin}, {Wade}, {Wade}, {Wade},
  {Walet}, {Walker}, {Wallace}, {Walsh}, {Wang}, {Wang}, {Wang}, {Wang},
  {Wang}, {Ward}, {Warner}, {Was}, {Watchi}, {Weaver}, {Wei}, {Weinert},
  {Weinstein}, {Weiss}, {Wen}, {Wessel}, {We{\ss}els}, {Westerweck},
  {Westphal}, {Wette}, {Whelan}, {Whitcomb}, {Whiting}, {Whittle}, {Wilken},
  {Williams}, {Williams}, {Williamson}, {Willis}, {Willke}, {Wimmer},
  {Winkler}, {Wipf}, {Wittel}, {Woan}, {Woehler}, {Wofford}, {Wong}, {Worden},
  {Wright}, {Wu}, {Wysocki}, {Xiao}, {Yamamoto}, {Yancey}, {Yang}, {Yap},
  {Yazback}, {Yu}, {Yu}, {Yvert}, {Zadro{\.Z}ny}, {Zanolin}, {Zelenova},
  {Zendri}, {Zevin}, {Zhang}, {Zhang}, {Zhang}, {Zhang}, {Zhao}, {Zhou},
  {Zhou}, {Zhu}, {Zhu}, {Zimmerman}, {Zucker}, {Zweizig}, {LIGO Scientific
  Collaboration}, \& {Virgo Collaboration}}]{abb17}
{Abbott}, B.~P., {Abbott}, R., {Abbott}, T.~D., {et~al.} 2017, \prl, 119,
  161101, \dodoi{10.1103/PhysRevLett.119.161101}

\bibitem[{{Abdurro'uf} {et~al.}(2022){Abdurro'uf}, {Accetta}, {Aerts}, {Silva
  Aguirre}, {Ahumada}, {Ajgaonkar}, {Filiz Ak}, {Alam}, {Allende Prieto},
  {Almeida}, {Anders}, {Anderson}, {Andrews}, {Anguiano}, {Aquino-Ort{\'\i}z},
  {Arag{\'o}n-Salamanca}, {Argudo-Fern{\'a}ndez}, {Ata}, {Aubert},
  {Avila-Reese}, {Badenes}, {Barb{\'a}}, {Barger}, {Barrera-Ballesteros},
  {Beaton}, {Beers}, {Belfiore}, {Bender}, {Bernardi}, {Bershady}, {Beutler},
  {Bidin}, {Bird}, {Bizyaev}, {Blanc}, {Blanton}, {Boardman}, {Bolton},
  {Boquien}, {Borissova}, {Bovy}, {Brandt}, {Brown}, {Brownstein}, {Brusa},
  {Buchner}, {Bundy}, {Burchett}, {Bureau}, {Burgasser}, {Cabang}, {Campbell},
  {Cappellari}, {Carlberg}, {Wanderley}, {Carrera}, {Cash}, {Chen}, {Chen},
  {Cherinka}, {Chiappini}, {Choi}, {Chojnowski}, {Chung}, {Clerc}, {Cohen},
  {Comerford}, {Comparat}, {da Costa}, {Covey}, {Crane}, {Cruz-Gonzalez},
  {Culhane}, {Cunha}, {Dai}, {Damke}, {Darling}, {Davidson}, {Davies},
  {Dawson}, {De Lee}, {Diamond-Stanic}, {Cano-D{\'\i}az}, {S{\'a}nchez},
  {Donor}, {Duckworth}, {Dwelly}, {Eisenstein}, {Elsworth}, {Emsellem},
  {Eracleous}, {Escoffier}, {Fan}, {Farr}, {Feng}, {Fern{\'a}ndez-Trincado},
  {Feuillet}, {Filipp}, {Fillingham}, {Frinchaboy}, {Fromenteau}, {Galbany},
  {Garc{\'\i}a}, {Garc{\'\i}a-Hern{\'a}ndez}, {Ge}, {Geisler}, {Gelfand},
  {G{\'e}ron}, {Gibson}, {Goddy}, {Godoy-Rivera}, {Grabowski}, {Green},
  {Greener}, {Grier}, {Griffith}, {Guo}, {Guy}, {Hadjara}, {Harding},
  {Hasselquist}, {Hayes}, {Hearty}, {Hern{\'a}ndez}, {Hill}, {Hogg},
  {Holtzman}, {Horta}, {Hsieh}, {Hsu}, {Hsu}, {Huber}, {Huertas-Company},
  {Hutchinson}, {Hwang}, {Ibarra-Medel}, {Chitham}, {Ilha}, {Imig}, {Jaekle},
  {Jayasinghe}, {Ji}, {Johnson}, {Jones}, {J{\"o}nsson}, {Katkov}, {Khalatyan},
  {Kinemuchi}, {Kisku}, {Knapen}, {Kneib}, {Kollmeier}, {Kong}, {Kounkel},
  {Kreckel}, {Krishnarao}, {Lacerna}, {Lane}, {Langgin}, {Lavender}, {Law},
  {Lazarz}, {Leung}, {Leung}, {Lewis}, {Li}, {Li}, {Lian}, {Liang}, {Lin},
  {Lin}, {Lin}, {Lintott}, {Long}, {Longa-Pe{\~n}a}, {L{\'o}pez-Cob{\'a}},
  {Lu}, {Lundgren}, {Luo}, {Mackereth}, {de la Macorra}, {Mahadevan},
  {Majewski}, {Manchado}, {Mandeville}, {Maraston}, {Margalef-Bentabol},
  {Masseron}, {Masters}, {Mathur}, {McDermid}, {Mckay}, {Merloni},
  {Merrifield}, {Meszaros}, {Miglio}, {Di Mille}, {Minniti}, {Minsley},
  {Monachesi}, {Moon}, {Mosser}, {Mulchaey}, {Muna}, {Mu{\~n}oz}, {Myers},
  {Myers}, {Nadathur}, {Nair}, {Nandra}, {Neumann}, {Newman}, {Nidever},
  {Nikakhtar}, {Nitschelm}, {O'Connell}, {Garma-Oehmichen}, {Luan Souza de
  Oliveira}, {Olney}, {Oravetz}, {Ortigoza-Urdaneta}, {Osorio}, {Otter},
  {Pace}, {Padilla}, {Pan}, {Pan}, {Parikh}, {Parker}, {Peirani}, {Pe{\~n}a
  Ram{\'\i}rez}, {Penny}, {Percival}, {Perez-Fournon}, {Pinsonneault},
  {Poidevin}, {Poovelil}, {Price-Whelan}, {B{\'a}rbara de Andrade Queiroz},
  {Raddick}, {Ray}, {Rembold}, {Riddle}, {Riffel}, {Riffel}, {Rix}, {Robin},
  {Rodr{\'\i}guez-Puebla}, {Roman-Lopes}, {Rom{\'a}n-Z{\'u}{\~n}iga}, {Rose},
  {Ross}, {Rossi}, {Rubin}, {Salvato}, {S{\'a}nchez}, {S{\'a}nchez-Gallego},
  {Sanderson}, {Santana Rojas}, {Sarceno}, {Sarmiento}, {Sayres}, {Sazonova},
  {Schaefer}, {Schiavon}, {Schlegel}, {Schneider}, {Schultheis}, {Schwope},
  {Serenelli}, {Serna}, {Shao}, {Shapiro}, {Sharma}, {Shen}, {Shetrone}, {Shu},
  {Simon}, {Skrutskie}, {Smethurst}, {Smith}, {Sobeck}, {Spoo}, {Sprague},
  {Stark}, {Stassun}, {Steinmetz}, {Stello}, {Stone-Martinez},
  {Storchi-Bergmann}, {Stringfellow}, {Stutz}, {Su}, {Taghizadeh-Popp},
  {Talbot}, {Tayar}, {Telles}, {Teske}, {Thakar}, {Theissen}, {Tkachenko},
  {Thomas}, {Tojeiro}, {Hernandez Toledo}, {Troup}, {Trump}, {Trussler},
  {Turner}, {Tuttle}, {Unda-Sanzana}, {V{\'a}zquez-Mata}, {Valentini},
  {Valenzuela}, {Vargas-Gonz{\'a}lez}, {Vargas-Maga{\~n}a}, {Alfaro},
  {Villanova}, {Vincenzo}, {Wake}, {Warfield}, {Washington}, {Weaver},
  {Weijmans}, {Weinberg}, {Weiss}, {Westfall}, {Wild}, {Wilde}, {Wilson},
  {Wilson}, {Wilson}, {Wolf}, {Wood-Vasey}, {Yan}, {Zamora}, {Zasowski},
  {Zhang}, {Zhao}, {Zheng}, {Zheng}, \& {Zhu}}]{abd22}
{Abdurro'uf}, {Accetta}, K., {Aerts}, C., {et~al.} 2022, \apjs, 259, 35,
  \dodoi{10.3847/1538-4365/ac4414}

\bibitem[{{Allende Prieto} {et~al.}(2006){Allende Prieto}, {Beers}, {Wilhelm},
  {Newberg}, {Rockosi}, {Yanny}, \& {Lee}}]{all06}
{Allende Prieto}, C., {Beers}, T.~C., {Wilhelm}, R., {et~al.} 2006, \apj, 636,
  804, \dodoi{10.1086/498131}

\bibitem[{{Alvarez} \& {Plez}(1998)}]{alv98}
{Alvarez}, R., \& {Plez}, B. 1998, \aap, 330, 1109.
\newblock \doarXiv{astro-ph/9710157}

\bibitem[{{Arcavi} {et~al.}(2017){Arcavi}, {Hosseinzadeh}, {Howell}, {McCully},
  {Poznanski}, {Kasen}, {Barnes}, {Zaltzman}, {Vasylyev}, {Maoz}, \&
  {Valenti}}]{arc17}
{Arcavi}, I., {Hosseinzadeh}, G., {Howell}, D.~A., {et~al.} 2017, \nat, 551,
  64, \dodoi{10.1038/nature24291}

\bibitem[{{Arenou} {et~al.}(2018){Arenou}, {Luri}, {Babusiaux}, {Fabricius},
  {Helmi}, {Muraveva}, {Robin}, {Spoto}, {Vallenari}, {Antoja},
  {Cantat-Gaudin}, {Jordi}, {Leclerc}, {Reyl{\'e}}, {Romero-G{\'o}mez}, {Shih},
  {Soria}, {Barache}, {Bossini}, {Bragaglia}, {Breddels}, {Fabrizio},
  {Lambert}, {Marrese}, {Massari}, {Moitinho}, {Robichon}, {Ruiz-Dern},
  {Sordo}, {Veljanoski}, {Eyer}, {Jasniewicz}, {Pancino}, {Soubiran}, {Spagna},
  {Tanga}, {Turon}, \& {Zurbach}}]{are18}
{Arenou}, F., {Luri}, X., {Babusiaux}, C., {et~al.} 2018, \aap, 616, A17,
  \dodoi{10.1051/0004-6361/201833234}

\bibitem[{{Astropy Collaboration} {et~al.}(2013){Astropy Collaboration},
  {Robitaille}, {Tollerud}, {Greenfield}, {Droettboom}, {Bray}, {Aldcroft},
  {Davis}, {Ginsburg}, {Price-Whelan}, {Kerzendorf}, {Conley}, {Crighton},
  {Barbary}, {Muna}, {Ferguson}, {Grollier}, {Parikh}, {Nair}, {Unther},
  {Deil}, {Woillez}, {Conseil}, {Kramer}, {Turner}, {Singer}, {Fox}, {Weaver},
  {Zabalza}, {Edwards}, {Azalee Bostroem}, {Burke}, {Casey}, {Crawford},
  {Dencheva}, {Ely}, {Jenness}, {Labrie}, {Lim}, {Pierfederici}, {Pontzen},
  {Ptak}, {Refsdal}, {Servillat}, \& {Streicher}}]{astropy13}
{Astropy Collaboration}, {Robitaille}, T.~P., {Tollerud}, E.~J., {et~al.} 2013,
  \aap, 558, A33, \dodoi{10.1051/0004-6361/201322068}

\bibitem[{{Astropy Collaboration} {et~al.}(2018){Astropy Collaboration},
  {Price-Whelan}, {Sip{\H{o}}cz}, {G{\"u}nther}, {Lim}, {Crawford}, {Conseil},
  {Shupe}, {Craig}, {Dencheva}, {Ginsburg}, {VanderPlas}, {Bradley},
  {P{\'e}rez-Su{\'a}rez}, {de Val-Borro}, {Aldcroft}, {Cruz}, {Robitaille},
  {Tollerud}, {Ardelean}, {Babej}, {Bach}, {Bachetti}, {Bakanov}, {Bamford},
  {Barentsen}, {Barmby}, {Baumbach}, {Berry}, {Biscani}, {Boquien}, {Bostroem},
  {Bouma}, {Brammer}, {Bray}, {Breytenbach}, {Buddelmeijer}, {Burke},
  {Calderone}, {Cano Rodr{\'\i}guez}, {Cara}, {Cardoso}, {Cheedella}, {Copin},
  {Corrales}, {Crichton}, {D'Avella}, {Deil}, {Depagne}, {Dietrich}, {Donath},
  {Droettboom}, {Earl}, {Erben}, {Fabbro}, {Ferreira}, {Finethy}, {Fox},
  {Garrison}, {Gibbons}, {Goldstein}, {Gommers}, {Greco}, {Greenfield},
  {Groener}, {Grollier}, {Hagen}, {Hirst}, {Homeier}, {Horton}, {Hosseinzadeh},
  {Hu}, {Hunkeler}, {Ivezi{\'c}}, {Jain}, {Jenness}, {Kanarek}, {Kendrew},
  {Kern}, {Kerzendorf}, {Khvalko}, {King}, {Kirkby}, {Kulkarni}, {Kumar},
  {Lee}, {Lenz}, {Littlefair}, {Ma}, {Macleod}, {Mastropietro}, {McCully},
  {Montagnac}, {Morris}, {Mueller}, {Mumford}, {Muna}, {Murphy}, {Nelson},
  {Nguyen}, {Ninan}, {N{\"o}the}, {Ogaz}, {Oh}, {Parejko}, {Parley}, {Pascual},
  {Patil}, {Patil}, {Plunkett}, {Prochaska}, {Rastogi}, {Reddy Janga},
  {Sabater}, {Sakurikar}, {Seifert}, {Sherbert}, {Sherwood-Taylor}, {Shih},
  {Sick}, {Silbiger}, {Singanamalla}, {Singer}, {Sladen}, {Sooley},
  {Sornarajah}, {Streicher}, {Teuben}, {Thomas}, {Tremblay}, {Turner},
  {Terr{\'o}n}, {van Kerkwijk}, {de la Vega}, {Watkins}, {Weaver}, {Whitmore},
  {Woillez}, {Zabalza}, \& {Astropy Contributors}}]{astropy18}
{Astropy Collaboration}, {Price-Whelan}, A.~M., {Sip{\H{o}}cz}, B.~M., {et~al.}
  2018, \aj, 156, 123, \dodoi{10.3847/1538-3881/aabc4f}

\bibitem[{{Bailer-Jones} {et~al.}(2021){Bailer-Jones}, {Rybizki}, {Fouesneau},
  {Demleitner}, \& {Andrae}}]{bai21}
{Bailer-Jones}, C.~A.~L., {Rybizki}, J., {Fouesneau}, M., {Demleitner}, M., \&
  {Andrae}, R. 2021, \aj, 161, 147, \dodoi{10.3847/1538-3881/abd806}

\bibitem[{{Bailer-Jones} {et~al.}(2018){Bailer-Jones}, {Rybizki}, {Fouesneau},
  {Mantelet}, \& {Andrae}}]{bai18}
{Bailer-Jones}, C.~A.~L., {Rybizki}, J., {Fouesneau}, M., {Mantelet}, G., \&
  {Andrae}, R. 2018, \aj, 156, 58, \dodoi{10.3847/1538-3881/aacb21}

\bibitem[{{Barklem} {et~al.}(2005){Barklem}, {Christlieb}, {Beers}, {Hill},
  {Bessell}, {Holmberg}, {Marsteller}, {Rossi}, {Zickgraf}, \&
  {Reimers}}]{bar05}
{Barklem}, P.~S., {Christlieb}, N., {Beers}, T.~C., {et~al.} 2005, \aap, 439,
  129, \dodoi{10.1051/0004-6361:20052967}

\bibitem[{{Bastian} \& {Lardo}(2018)}]{bas18}
{Bastian}, N., \& {Lardo}, C. 2018, \araa, 56, 83,
  \dodoi{10.1146/annurev-astro-081817-051839}

\bibitem[{{Beaton} {et~al.}(2021){Beaton}, {Oelkers}, {Hayes}, {Covey},
  {Chojnowski}, {De Lee}, {Sobeck}, {Majewski}, {Cohen},
  {Fern{\'a}ndez-Trincado}, {Longa-Pe{\~n}a}, {O'Connell}, {Santana},
  {Stringfellow}, {Zasowski}, {Aerts}, {Anguiano}, {Bender}, {Ca{\~n}as},
  {Cunha}, {Donor}, {Fleming}, {Frinchaboy}, {Feuillet}, {Harding},
  {Hasselquist}, {Holtzman}, {Johnson}, {Kollmeier}, {Kounkel}, {Mahadevan},
  {Price-Whelan}, {Rojas-Arriagada}, {Rom{\'a}n-Z{\'u}{\~n}iga}, {Schlafly},
  {Schultheis}, {Shetrone}, {Simon}, {Stassun}, {Stutz}, {Tayar}, {Teske},
  {Tkachenko}, {Troup}, {Albareti}, {Bizyaev}, {Bovy}, {Burgasser}, {Comparat},
  {Downes}, {Geisler}, {Inno}, {Manchado}, {Ness}, {Pinsonneault}, {Prada},
  {Roman-Lopes}, {Simonian}, {Smith}, {Yan}, \& {Zamora}}]{bea21}
{Beaton}, R.~L., {Oelkers}, R.~J., {Hayes}, C.~R., {et~al.} 2021, \aj, 162,
  302, \dodoi{10.3847/1538-3881/ac260c}

\bibitem[{{Beers} \& {Christlieb}(2005)}]{bee05}
{Beers}, T.~C., \& {Christlieb}, N. 2005, \araa, 43, 531,
  \dodoi{10.1146/annurev.astro.42.053102.134057}

\bibitem[{{Bekki} \& {Tsujimoto}(2012)}]{bek12}
{Bekki}, K., \& {Tsujimoto}, T. 2012, \apj, 761, 180,
  \dodoi{10.1088/0004-637X/761/2/180}

\bibitem[{{Blanton} {et~al.}(2017){Blanton}, {Bershady}, {Abolfathi},
  {Albareti}, {Allende Prieto}, {Almeida}, {Alonso-Garc{\'\i}a}, {Anders},
  {Anderson}, {Andrews}, {Aquino-Ort{\'\i}z}, {Arag{\'o}n-Salamanca},
  {Argudo-Fern{\'a}ndez}, {Armengaud}, {Aubourg}, {Avila-Reese}, {Badenes},
  {Bailey}, {Barger}, {Barrera-Ballesteros}, {Bartosz}, {Bates}, {Baumgarten},
  {Bautista}, {Beaton}, {Beers}, {Belfiore}, {Bender}, {Berlind}, {Bernardi},
  {Beutler}, {Bird}, {Bizyaev}, {Blanc}, {Blomqvist}, {Bolton}, {Boquien},
  {Borissova}, {van den Bosch}, {Bovy}, {Brandt}, {Brinkmann}, {Brownstein},
  {Bundy}, {Burgasser}, {Burtin}, {Busca}, {Cappellari}, {Delgado Carigi},
  {Carlberg}, {Carnero Rosell}, {Carrera}, {Chanover}, {Cherinka}, {Cheung},
  {G{\'o}mez Maqueo Chew}, {Chiappini}, {Choi}, {Chojnowski}, {Chuang},
  {Chung}, {Cirolini}, {Clerc}, {Cohen}, {Comparat}, {da Costa}, {Cousinou},
  {Covey}, {Crane}, {Croft}, {Cruz-Gonzalez}, {Garrido Cuadra}, {Cunha},
  {Damke}, {Darling}, {Davies}, {Dawson}, {de la Macorra}, {Dell'Agli}, {De
  Lee}, {Delubac}, {Di Mille}, {Diamond-Stanic}, {Cano-D{\'\i}az}, {Donor},
  {Downes}, {Drory}, {du Mas des Bourboux}, {Duckworth}, {Dwelly}, {Dyer},
  {Ebelke}, {Eigenbrot}, {Eisenstein}, {Emsellem}, {Eracleous}, {Escoffier},
  {Evans}, {Fan}, {Fern{\'a}ndez-Alvar}, {Fernandez-Trincado}, {Feuillet},
  {Finoguenov}, {Fleming}, {Font-Ribera}, {Fredrickson}, {Freischlad},
  {Frinchaboy}, {Fuentes}, {Galbany}, {Garcia-Dias},
  {Garc{\'\i}a-Hern{\'a}ndez}, {Gaulme}, {Geisler}, {Gelfand},
  {Gil-Mar{\'\i}n}, {Gillespie}, {Goddard}, {Gonzalez-Perez}, {Grabowski},
  {Green}, {Grier}, {Gunn}, {Guo}, {Guy}, {Hagen}, {Hahn}, {Hall}, {Harding},
  {Hasselquist}, {Hawley}, {Hearty}, {Gonzalez Hern{\'a}ndez}, {Ho}, {Hogg},
  {Holley-Bockelmann}, {Holtzman}, {Holzer}, {Huehnerhoff}, {Hutchinson},
  {Hwang}, {Ibarra-Medel}, {da Silva Ilha}, {Ivans}, {Ivory}, {Jackson},
  {Jensen}, {Johnson}, {Jones}, {J{\"o}nsson}, {Jullo}, {Kamble}, {Kinemuchi},
  {Kirkby}, {Kitaura}, {Klaene}, {Knapp}, {Kneib}, {Kollmeier}, {Lacerna},
  {Lane}, {Lang}, {Law}, {Lazarz}, {Lee}, {Le Goff}, {Liang}, {Li}, {Li},
  {Lian}, {Lima}, {Lin}, {Lin}, {Bertran de Lis}, {Liu}, {de Icaza Lizaola},
  {Long}, {Lucatello}, {Lundgren}, {MacDonald}, {Deconto Machado}, {MacLeod},
  {Mahadevan}, {Geimba Maia}, {Maiolino}, {Majewski}, {Malanushenko},
  {Malanushenko}, {Manchado}, {Mao}, {Maraston}, {Marques-Chaves}, {Masseron},
  {Masters}, {McBride}, {McDermid}, {McGrath}, {McGreer}, {Medina Pe{\~n}a},
  {Melendez}, {Merloni}, {Merrifield}, {Meszaros}, {Meza}, {Minchev},
  {Minniti}, {Miyaji}, {More}, {Mulchaey}, {M{\"u}ller-S{\'a}nchez}, {Muna},
  {Munoz}, {Myers}, {Nair}, {Nandra}, {Correa do Nascimento}, {Negrete},
  {Ness}, {Newman}, {Nichol}, {Nidever}, {Nitschelm}, {Ntelis}, {O'Connell},
  {Oelkers}, {Oravetz}, {Oravetz}, {Pace}, {Padilla}, {Palanque-Delabrouille},
  {Alonso Palicio}, {Pan}, {Parejko}, {Parikh}, {P{\^a}ris}, {Park}, {Patten},
  {Peirani}, {Pellejero-Ibanez}, {Penny}, {Percival}, {Perez-Fournon},
  {Petitjean}, {Pieri}, {Pinsonneault}, {Pisani}, {Poleski}, {Prada},
  {Prakash}, {Queiroz}, {Raddick}, {Raichoor}, {Barboza Rembold}, {Richstein},
  {Riffel}, {Riffel}, {Rix}, {Robin}, {Rockosi}, {Rodr{\'\i}guez-Torres},
  {Roman-Lopes}, {Rom{\'a}n-Z{\'u}{\~n}iga}, {Rosado}, {Ross}, {Rossi}, {Ruan},
  {Ruggeri}, {Rykoff}, {Salazar-Albornoz}, {Salvato}, {S{\'a}nchez}, {Aguado},
  {S{\'a}nchez-Gallego}, {Santana}, {Santiago}, {Sayres}, {Schiavon}, {da Silva
  Schimoia}, {Schlafly}, {Schlegel}, {Schneider}, {Schultheis}, {Schuster},
  {Schwope}, {Seo}, {Shao}, {Shen}, {Shetrone}, {Shull}, {Simon}, {Skinner},
  {Skrutskie}, {Slosar}, {Smith}, {Sobeck}, {Sobreira}, {Somers}, {Souto},
  {Stark}, {Stassun}, {Stauffer}, {Steinmetz}, {Storchi-Bergmann},
  {Streblyanska}, {Stringfellow}, {Su{\'a}rez}, {Sun}, {Suzuki}, {Szigeti},
  {Taghizadeh-Popp}, {Tang}, {Tao}, {Tayar}, {Tembe}, {Teske}, {Thakar},
  {Thomas}, {Thompson}, {Tinker}, {Tissera}, {Tojeiro}, {Hernandez Toledo}, {de
  la Torre}, {Tremonti}, {Troup}, {Valenzuela}, {Martinez Valpuesta},
  {Vargas-Gonz{\'a}lez}, {Vargas-Maga{\~n}a}, {Vazquez}, {Villanova}, {Vivek},
  {Vogt}, {Wake}, {Walterbos}, {Wang}, {Weaver}, {Weijmans}, {Weinberg},
  {Westfall}, {Whelan}, {Wild}, {Wilson}, {Wood-Vasey}, {Wylezalek}, {Xiao},
  {Yan}, {Yang}, {Ybarra}, {Y{\`e}che}, {Zakamska}, {Zamora}, {Zarrouk},
  {Zasowski}, {Zhang}, {Zhao}, {Zheng}, {Zheng}, {Zhou}, {Zhou}, {Zhu},
  {Zoccali}, \& {Zou}}]{bla17}
{Blanton}, M.~R., {Bershady}, M.~A., {Abolfathi}, B., {et~al.} 2017, \aj, 154,
  28, \dodoi{10.3847/1538-3881/aa7567}

\bibitem[{{Boos} {et~al.}(2021){Boos}, {Townsley}, {Shen}, {Caldwell}, \&
  {Miles}}]{boo21}
{Boos}, S.~J., {Townsley}, D.~M., {Shen}, K.~J., {Caldwell}, S., \& {Miles},
  B.~J. 2021, \apj, 919, 126, \dodoi{10.3847/1538-4357/ac07a2}

\bibitem[{{Bowen} \& {Vaughan}(1973)}]{bow73}
{Bowen}, I.~S., \& {Vaughan}, A.~H., J. 1973, \ao, 12, 1430,
  \dodoi{10.1364/AO.12.001430}

\bibitem[{{Bravo} {et~al.}(2019){Bravo}, {Badenes}, \&
  {Mart{\'\i}nez-Rodr{\'\i}guez}}]{bra19}
{Bravo}, E., {Badenes}, C., \& {Mart{\'\i}nez-Rodr{\'\i}guez}, H. 2019, \mnras,
  482, 4346, \dodoi{10.1093/mnras/sty2951}

\bibitem[{{Calura} {et~al.}(2019){Calura}, {D'Ercole}, {Vesperini}, {Vanzella},
  \& {Sollima}}]{cal19}
{Calura}, F., {D'Ercole}, A., {Vesperini}, E., {Vanzella}, E., \& {Sollima}, A.
  2019, \mnras, 489, 3269, \dodoi{10.1093/mnras/stz2055}

\bibitem[{{Carney} {et~al.}(2003){Carney}, {Latham}, {Stefanik}, {Laird}, \&
  {Morse}}]{car03}
{Carney}, B.~W., {Latham}, D.~W., {Stefanik}, R.~P., {Laird}, J.~B., \&
  {Morse}, J.~A. 2003, \aj, 125, 293, \dodoi{10.1086/345386}

\bibitem[{{Carretta} {et~al.}(2009){Carretta}, {Bragaglia}, {Gratton},
  {Lucatello}, {Catanzaro}, {Leone}, {Bellazzini}, {Claudi}, {D'Orazi},
  {Momany}, {Ortolani}, {Pancino}, {Piotto}, {Recio-Blanco}, \&
  {Sabbi}}]{car09}
{Carretta}, E., {Bragaglia}, A., {Gratton}, R.~G., {et~al.} 2009, \aap, 505,
  117, \dodoi{10.1051/0004-6361/200912096}

\bibitem[{{Choi} {et~al.}(2016){Choi}, {Dotter}, {Conroy}, {Cantiello},
  {Paxton}, \& {Johnson}}]{cho16}
{Choi}, J., {Dotter}, A., {Conroy}, C., {et~al.} 2016, \apj, 823, 102,
  \dodoi{10.3847/0004-637X/823/2/102}

\bibitem[{{Cohen}(1978)}]{coh78}
{Cohen}, J.~G. 1978, \apj, 223, 487, \dodoi{10.1086/156284}

\bibitem[{{Cohen}(1979)}]{coh79}
---. 1979, \apj, 231, 751, \dodoi{10.1086/157241}

\bibitem[{{Cohen}(1980)}]{coh80}
---. 1980, \apj, 241, 981, \dodoi{10.1086/158412}

\bibitem[{{Cohen}(1981)}]{coh81}
---. 1981, \apj, 247, 869, \dodoi{10.1086/159097}

\bibitem[{{Conroy}(2012)}]{con12}
{Conroy}, C. 2012, \apj, 758, 21, \dodoi{10.1088/0004-637X/758/1/21}

\bibitem[{{Cunha} {et~al.}(2015){Cunha}, {Smith}, {Johnson}, {Bergemann},
  {M{\'e}sz{\'a}ros}, {Shetrone}, {Souto}, {Allende Prieto}, {Schiavon},
  {Frinchaboy}, {Zasowski}, {Bizyaev}, {Holtzman}, {Garc{\'\i}a P{\'e}rez},
  {Majewski}, {Nidever}, {Beers}, {Carrera}, {Geisler}, {Gunn}, {Hearty},
  {Ivans}, {Martell}, {Pinsonneault}, {Schneider}, {Sobeck}, {Stello},
  {Stassun}, {Skrutskie}, \& {Wilson}}]{cun15}
{Cunha}, K., {Smith}, V.~V., {Johnson}, J.~A., {et~al.} 2015, \apjl, 798, L41,
  \dodoi{10.1088/2041-8205/798/2/L41}

\bibitem[{{D'Ercole} {et~al.}(2008){D'Ercole}, {Vesperini}, {D'Antona},
  {McMillan}, \& {Recchi}}]{der08}
{D'Ercole}, A., {Vesperini}, E., {D'Antona}, F., {McMillan}, S. L.~W., \&
  {Recchi}, S. 2008, \mnras, 391, 825, \dodoi{10.1111/j.1365-2966.2008.13915.x}

\bibitem[{{Dotter}(2016)}]{dot16}
{Dotter}, A. 2016, \apjs, 222, 8, \dodoi{10.3847/0067-0049/222/1/8}

\bibitem[{{Eisenstein} {et~al.}(2011){Eisenstein}, {Weinberg}, {Agol},
  {Aihara}, {Allende Prieto}, {Anderson}, {Arns}, {Aubourg}, {Bailey},
  {Balbinot}, {Barkhouser}, {Beers}, {Berlind}, {Bickerton}, {Bizyaev},
  {Blanton}, {Bochanski}, {Bolton}, {Bosman}, {Bovy}, {Brandt}, {Breslauer},
  {Brewington}, {Brinkmann}, {Brown}, {Brownstein}, {Burger}, {Busca},
  {Campbell}, {Cargile}, {Carithers}, {Carlberg}, {Carr}, {Chang}, {Chen},
  {Chiappini}, {Comparat}, {Connolly}, {Cortes}, {Croft}, {Cunha}, {da Costa},
  {Davenport}, {Dawson}, {De Lee}, {Porto de Mello}, {de Simoni}, {Dean},
  {Dhital}, {Ealet}, {Ebelke}, {Edmondson}, {Eiting}, {Escoffier}, {Esposito},
  {Evans}, {Fan}, {Femen{\'\i}a Castell{\'a}}, {Dutra Ferreira}, {Fitzgerald},
  {Fleming}, {Font-Ribera}, {Ford}, {Frinchaboy}, {Garc{\'\i}a P{\'e}rez},
  {Gaudi}, {Ge}, {Ghezzi}, {Gillespie}, {Gilmore}, {Girardi}, {Gott}, {Gould},
  {Grebel}, {Gunn}, {Hamilton}, {Harding}, {Harris}, {Hawley}, {Hearty},
  {Hennawi}, {Gonz{\'a}lez Hern{\'a}ndez}, {Ho}, {Hogg}, {Holtzman},
  {Honscheid}, {Inada}, {Ivans}, {Jiang}, {Jiang}, {Johnson}, {Jordan},
  {Jordan}, {Kauffmann}, {Kazin}, {Kirkby}, {Klaene}, {Knapp}, {Kneib},
  {Kochanek}, {Koesterke}, {Kollmeier}, {Kron}, {Lampeitl}, {Lang}, {Lawler},
  {Le Goff}, {Lee}, {Lee}, {Leisenring}, {Lin}, {Liu}, {Long}, {Loomis},
  {Lucatello}, {Lundgren}, {Lupton}, {Ma}, {Ma}, {MacDonald}, {Mack},
  {Mahadevan}, {Maia}, {Majewski}, {Makler}, {Malanushenko}, {Malanushenko},
  {Mandelbaum}, {Maraston}, {Margala}, {Maseman}, {Masters}, {McBride},
  {McDonald}, {McGreer}, {McMahon}, {Mena Requejo}, {M{\'e}nard},
  {Miralda-Escud{\'e}}, {Morrison}, {Mullally}, {Muna}, {Murayama}, {Myers},
  {Naugle}, {Neto}, {Nguyen}, {Nichol}, {Nidever}, {O'Connell}, {Ogando},
  {Olmstead}, {Oravetz}, {Padmanabhan}, {Paegert}, {Palanque-Delabrouille},
  {Pan}, {Pandey}, {Parejko}, {P{\^a}ris}, {Pellegrini}, {Pepper}, {Percival},
  {Petitjean}, {Pfaffenberger}, {Pforr}, {Phleps}, {Pichon}, {Pieri}, {Prada},
  {Price-Whelan}, {Raddick}, {Ramos}, {Reid}, {Reyle}, {Rich}, {Richards},
  {Rieke}, {Rieke}, {Rix}, {Robin}, {Rocha-Pinto}, {Rockosi}, {Roe},
  {Rollinde}, {Ross}, {Ross}, {Rossetto}, {S{\'a}nchez}, {Santiago}, {Sayres},
  {Schiavon}, {Schlegel}, {Schlesinger}, {Schmidt}, {Schneider}, {Sellgren},
  {Shelden}, {Sheldon}, {Shetrone}, {Shu}, {Silverman}, {Simmerer}, {Simmons},
  {Sivarani}, {Skrutskie}, {Slosar}, {Smee}, {Smith}, {Snedden}, {Stassun},
  {Steele}, {Steinmetz}, {Stockett}, {Stollberg}, {Strauss}, {Szalay},
  {Tanaka}, {Thakar}, {Thomas}, {Tinker}, {Tofflemire}, {Tojeiro}, {Tremonti},
  {Vargas Maga{\~n}a}, {Verde}, {Vogt}, {Wake}, {Wan}, {Wang}, {Weaver},
  {White}, {White}, {Wilson}, {Wisniewski}, {Wood-Vasey}, {Yanny}, {Yasuda},
  {Y{\`e}che}, {York}, {Young}, {Zasowski}, {Zehavi}, \& {Zhao}}]{eis11}
{Eisenstein}, D.~J., {Weinberg}, D.~H., {Agol}, E., {et~al.} 2011, \aj, 142,
  72, \dodoi{10.1088/0004-6256/142/3/72}

\bibitem[{{Evans} {et~al.}(2018){Evans}, {Riello}, {De Angeli}, {Carrasco},
  {Montegriffo}, {Fabricius}, {Jordi}, {Palaversa}, {Diener}, {Busso},
  {Cacciari}, {van Leeuwen}, {Burgess}, {Davidson}, {Harrison}, {Hodgkin},
  {Pancino}, {Richards}, {Altavilla}, {Balaguer-N{\'u}{\~n}ez}, {Barstow},
  {Bellazzini}, {Brown}, {Castellani}, {Cocozza}, {De Luise}, {Delgado},
  {Ducourant}, {Galleti}, {Gilmore}, {Giuffrida}, {Holl}, {Kewley}, {Koposov},
  {Marinoni}, {Marrese}, {Osborne}, {Piersimoni}, {Portell}, {Pulone},
  {Ragaini}, {Sanna}, {Terrett}, {Walton}, {Wevers}, \& {Wyrzykowski}}]{eva18}
{Evans}, D.~W., {Riello}, M., {De Angeli}, F., {et~al.} 2018, \aap, 616, A4,
  \dodoi{10.1051/0004-6361/201832756}

\bibitem[{{Ezzeddine} {et~al.}(2020){Ezzeddine}, {Rasmussen}, {Frebel},
  {Chiti}, {Hinojisa}, {Placco}, {Ji}, {Beers}, {Hansen}, {Roederer}, {Sakari},
  \& {Melendez}}]{ezz20}
{Ezzeddine}, R., {Rasmussen}, K., {Frebel}, A., {et~al.} 2020, \apj, 898, 150,
  \dodoi{10.3847/1538-4357/ab9d1a}

\bibitem[{{Fabricius} {et~al.}(2021){Fabricius}, {Luri}, {Arenou}, {Babusiaux},
  {Helmi}, {Muraveva}, {Reyl{\'e}}, {Spoto}, {Vallenari}, {Antoja}, {Balbinot},
  {Barache}, {Bauchet}, {Bragaglia}, {Busonero}, {Cantat-Gaudin}, {Carrasco},
  {Diakit{\'e}}, {Fabrizio}, {Figueras}, {Garcia-Gutierrez}, {Garofalo},
  {Jordi}, {Kervella}, {Khanna}, {Leclerc}, {Licata}, {Lambert}, {Marrese},
  {Masip}, {Ramos}, {Robichon}, {Robin}, {Romero-G{\'o}mez}, {Rubele}, \&
  {Weiler}}]{fab21}
{Fabricius}, C., {Luri}, X., {Arenou}, F., {et~al.} 2021, \aap, 649, A5,
  \dodoi{10.1051/0004-6361/202039834}

\bibitem[{{Fink} {et~al.}(2014){Fink}, {Kromer}, {Seitenzahl},
  {Ciaraldi-Schoolmann}, {R{\"o}pke}, {Sim}, {Pakmor}, {Ruiter}, \&
  {Hillebrandt}}]{fin14}
{Fink}, M., {Kromer}, M., {Seitenzahl}, I.~R., {et~al.} 2014, \mnras, 438,
  1762, \dodoi{10.1093/mnras/stt2315}

\bibitem[{{Gaia Collaboration} {et~al.}(2016){Gaia Collaboration}, {Prusti},
  {de Bruijne}, {Brown}, {Vallenari}, {Babusiaux}, {Bailer-Jones}, {Bastian},
  {Biermann}, {Evans}, {Eyer}, {Jansen}, {Jordi}, {Klioner}, {Lammers},
  {Lindegren}, {Luri}, {Mignard}, {Milligan}, {Panem}, {Poinsignon},
  {Pourbaix}, {Randich}, {Sarri}, {Sartoretti}, {Siddiqui}, {Soubiran},
  {Valette}, {van Leeuwen}, {Walton}, {Aerts}, {Arenou}, {Cropper}, {Drimmel},
  {H{\o}g}, {Katz}, {Lattanzi}, {O'Mullane}, {Grebel}, {Holland}, {Huc},
  {Passot}, {Bramante}, {Cacciari}, {Casta{\~n}eda}, {Chaoul}, {Cheek}, {De
  Angeli}, {Fabricius}, {Guerra}, {Hern{\'a}ndez}, {Jean-Antoine-Piccolo},
  {Masana}, {Messineo}, {Mowlavi}, {Nienartowicz}, {Ord{\'o}{\~n}ez-Blanco},
  {Panuzzo}, {Portell}, {Richards}, {Riello}, {Seabroke}, {Tanga},
  {Th{\'e}venin}, {Torra}, {Els}, {Gracia-Abril}, {Comoretto},
  {Garcia-Reinaldos}, {Lock}, {Mercier}, {Altmann}, {Andrae}, {Astraatmadja},
  {Bellas-Velidis}, {Benson}, {Berthier}, {Blomme}, {Busso}, {Carry},
  {Cellino}, {Clementini}, {Cowell}, {Creevey}, {Cuypers}, {Davidson}, {De
  Ridder}, {de Torres}, {Delchambre}, {Dell'Oro}, {Ducourant}, {Fr{\'e}mat},
  {Garc{\'\i}a-Torres}, {Gosset}, {Halbwachs}, {Hambly}, {Harrison}, {Hauser},
  {Hestroffer}, {Hodgkin}, {Huckle}, {Hutton}, {Jasniewicz}, {Jordan},
  {Kontizas}, {Korn}, {Lanzafame}, {Manteiga}, {Moitinho}, {Muinonen},
  {Osinde}, {Pancino}, {Pauwels}, {Petit}, {Recio-Blanco}, {Robin}, {Sarro},
  {Siopis}, {Smith}, {Smith}, {Sozzetti}, {Thuillot}, {van Reeven}, {Viala},
  {Abbas}, {Abreu Aramburu}, {Accart}, {Aguado}, {Allan}, {Allasia},
  {Altavilla}, {{\'A}lvarez}, {Alves}, {Anderson}, {Andrei}, {Anglada Varela},
  {Antiche}, {Antoja}, {Ant{\'o}n}, {Arcay}, {Atzei}, {Ayache}, {Bach},
  {Baker}, {Balaguer-N{\'u}{\~n}ez}, {Barache}, {Barata}, {Barbier}, {Barblan},
  {Baroni}, {Barrado y Navascu{\'e}s}, {Barros}, {Barstow}, {Becciani},
  {Bellazzini}, {Bellei}, {Bello Garc{\'\i}a}, {Belokurov}, {Bendjoya},
  {Berihuete}, {Bianchi}, {Bienaym{\'e}}, {Billebaud}, {Blagorodnova},
  {Blanco-Cuaresma}, {Boch}, {Bombrun}, {Borrachero}, {Bouquillon}, {Bourda},
  {Bouy}, {Bragaglia}, {Breddels}, {Brouillet}, {Br{\"u}semeister},
  {Bucciarelli}, {Budnik}, {Burgess}, {Burgon}, {Burlacu}, {Busonero}, {Buzzi},
  {Caffau}, {Cambras}, {Campbell}, {Cancelliere}, {Cantat-Gaudin}, {Carlucci},
  {Carrasco}, {Castellani}, {Charlot}, {Charnas}, {Charvet}, {Chassat},
  {Chiavassa}, {Clotet}, {Cocozza}, {Collins}, {Collins}, {Costigan}, {Crifo},
  {Cross}, {Crosta}, {Crowley}, {Dafonte}, {Damerdji}, {Dapergolas}, {David},
  {David}, {De Cat}, {de Felice}, {de Laverny}, {De Luise}, {De March}, {de
  Martino}, {de Souza}, {Debosscher}, {del Pozo}, {Delbo}, {Delgado},
  {Delgado}, {di Marco}, {Di Matteo}, {Diakite}, {Distefano}, {Dolding}, {Dos
  Anjos}, {Drazinos}, {Dur{\'a}n}, {Dzigan}, {Ecale}, {Edvardsson}, {Enke},
  {Erdmann}, {Escolar}, {Espina}, {Evans}, {Eynard Bontemps}, {Fabre},
  {Fabrizio}, {Faigler}, {Falc{\~a}o}, {Farr{\`a}s Casas}, {Faye}, {Federici},
  {Fedorets}, {Fern{\'a}ndez-Hern{\'a}ndez}, {Fernique}, {Fienga}, {Figueras},
  {Filippi}, {Findeisen}, {Fonti}, {Fouesneau}, {Fraile}, {Fraser}, {Fuchs},
  {Furnell}, {Gai}, {Galleti}, {Galluccio}, {Garabato}, {Garc{\'\i}a-Sedano},
  {Gar{\'e}}, {Garofalo}, {Garralda}, {Gavras}, {Gerssen}, {Geyer}, {Gilmore},
  {Girona}, {Giuffrida}, {Gomes}, {Gonz{\'a}lez-Marcos},
  {Gonz{\'a}lez-N{\'u}{\~n}ez}, {Gonz{\'a}lez-Vidal}, {Granvik}, {Guerrier},
  {Guillout}, {Guiraud}, {G{\'u}rpide}, {Guti{\'e}rrez-S{\'a}nchez}, {Guy},
  {Haigron}, {Hatzidimitriou}, {Haywood}, {Heiter}, {Helmi}, {Hobbs},
  {Hofmann}, {Holl}, {Holland}, {Hunt}, {Hypki}, {Icardi}, {Irwin}, {Jevardat
  de Fombelle}, {Jofr{\'e}}, {Jonker}, {Jorissen}, {Julbe}, {Karampelas},
  {Kochoska}, {Kohley}, {Kolenberg}, {Kontizas}, {Koposov}, {Kordopatis},
  {Koubsky}, {Kowalczyk}, {Krone-Martins}, {Kudryashova}, {Kull}, {Bachchan},
  {Lacoste-Seris}, {Lanza}, {Lavigne}, {Le Poncin-Lafitte}, {Lebreton},
  {Lebzelter}, {Leccia}, {Leclerc}, {Lecoeur-Taibi}, {Lemaitre}, {Lenhardt},
  {Leroux}, {Liao}, {Licata}, {Lindstr{\o}m}, {Lister}, {Livanou}, {Lobel},
  {L{\"o}ffler}, {L{\'o}pez}, {Lopez-Lozano}, {Lorenz}, {Loureiro},
  {MacDonald}, {Magalh{\~a}es Fernandes}, {Managau}, {Mann}, {Mantelet},
  {Marchal}, {Marchant}, {Marconi}, {Marie}, {Marinoni}, {Marrese},
  {Marschalk{\'o}}, {Marshall}, {Mart{\'\i}n-Fleitas}, {Martino}, {Mary},
  {Matijevi{\v{c}}}, {Mazeh}, {McMillan}, {Messina}, {Mestre}, {Michalik},
  {Millar}, {Miranda}, {Molina}, {Molinaro}, {Molinaro}, {Moln{\'a}r},
  {Moniez}, {Montegriffo}, {Monteiro}, {Mor}, {Mora}, {Morbidelli}, {Morel},
  {Morgenthaler}, {Morley}, {Morris}, {Mulone}, {Muraveva}, {Musella},
  {Narbonne}, {Nelemans}, {Nicastro}, {Noval}, {Ord{\'e}novic},
  {Ordieres-Mer{\'e}}, {Osborne}, {Pagani}, {Pagano}, {Pailler}, {Palacin},
  {Palaversa}, {Parsons}, {Paulsen}, {Pecoraro}, {Pedrosa}, {Pentik{\"a}inen},
  {Pereira}, {Pichon}, {Piersimoni}, {Pineau}, {Plachy}, {Plum}, {Poujoulet},
  {Pr{\v{s}}a}, {Pulone}, {Ragaini}, {Rago}, {Rambaux}, {Ramos-Lerate},
  {Ranalli}, {Rauw}, {Read}, {Regibo}, {Renk}, {Reyl{\'e}}, {Ribeiro},
  {Rimoldini}, {Ripepi}, {Riva}, {Rixon}, {Roelens}, {Romero-G{\'o}mez},
  {Rowell}, {Royer}, {Rudolph}, {Ruiz-Dern}, {Sadowski}, {Sagrist{\`a}
  Sell{\'e}s}, {Sahlmann}, {Salgado}, {Salguero}, {Sarasso}, {Savietto},
  {Schnorhk}, {Schultheis}, {Sciacca}, {Segol}, {Segovia}, {Segransan},
  {Serpell}, {Shih}, {Smareglia}, {Smart}, {Smith}, {Solano}, {Solitro},
  {Sordo}, {Soria Nieto}, {Souchay}, {Spagna}, {Spoto}, {Stampa}, {Steele},
  {Steidelm{\"u}ller}, {Stephenson}, {Stoev}, {Suess}, {S{\"u}veges}, {Surdej},
  {Szabados}, {Szegedi-Elek}, {Tapiador}, {Taris}, {Tauran}, {Taylor},
  {Teixeira}, {Terrett}, {Tingley}, {Trager}, {Turon}, {Ulla}, {Utrilla},
  {Valentini}, {van Elteren}, {Van Hemelryck}, {van Leeuwen}, {Varadi},
  {Vecchiato}, {Veljanoski}, {Via}, {Vicente}, {Vogt}, {Voss}, {Votruba},
  {Voutsinas}, {Walmsley}, {Weiler}, {Weingrill}, {Werner}, {Wevers},
  {Whitehead}, {Wyrzykowski}, {Yoldas}, {{\v{Z}}erjal}, {Zucker}, {Zurbach},
  {Zwitter}, {Alecu}, {Allen}, {Allende Prieto}, {Amorim},
  {Anglada-Escud{\'e}}, {Arsenijevic}, {Azaz}, {Balm}, {Beck}, {Bernstein},
  {Bigot}, {Bijaoui}, {Blasco}, {Bonfigli}, {Bono}, {Boudreault}, {Bressan},
  {Brown}, {Brunet}, {Bunclark}, {Buonanno}, {Butkevich}, {Carret}, {Carrion},
  {Chemin}, {Ch{\'e}reau}, {Corcione}, {Darmigny}, {de Boer}, {de Teodoro}, {de
  Zeeuw}, {Delle Luche}, {Domingues}, {Dubath}, {Fodor}, {Fr{\'e}zouls},
  {Fries}, {Fustes}, {Fyfe}, {Gallardo}, {Gallegos}, {Gardiol}, {Gebran},
  {Gomboc}, {G{\'o}mez}, {Grux}, {Gueguen}, {Heyrovsky}, {Hoar}, {Iannicola},
  {Isasi Parache}, {Janotto}, {Joliet}, {Jonckheere}, {Keil}, {Kim},
  {Klagyivik}, {Klar}, {Knude}, {Kochukhov}, {Kolka}, {Kos}, {Kutka}, {Lainey},
  {LeBouquin}, {Liu}, {Loreggia}, {Makarov}, {Marseille}, {Martayan},
  {Martinez-Rubi}, {Massart}, {Meynadier}, {Mignot}, {Munari}, {Nguyen},
  {Nordlander}, {Ocvirk}, {O'Flaherty}, {Olias Sanz}, {Ortiz}, {Osorio},
  {Oszkiewicz}, {Ouzounis}, {Palmer}, {Park}, {Pasquato}, {Peltzer}, {Peralta},
  {P{\'e}turaud}, {Pieniluoma}, {Pigozzi}, {Poels}, {Prat}, {Prod'homme},
  {Raison}, {Rebordao}, {Risquez}, {Rocca-Volmerange}, {Rosen}, {Ruiz-Fuertes},
  {Russo}, {Sembay}, {Serraller Vizcaino}, {Short}, {Siebert}, {Silva},
  {Sinachopoulos}, {Slezak}, {Soffel}, {Sosnowska}, {Strai{\v{z}}ys}, {ter
  Linden}, {Terrell}, {Theil}, {Tiede}, {Troisi}, {Tsalmantza}, {Tur},
  {Vaccari}, {Vachier}, {Valles}, {Van Hamme}, {Veltz}, {Virtanen}, {Wallut},
  {Wichmann}, {Wilkinson}, {Ziaeepour}, \& {Zschocke}}]{gaia16}
{Gaia Collaboration}, {Prusti}, T., {de Bruijne}, J.~H.~J., {et~al.} 2016,
  \aap, 595, A1, \dodoi{10.1051/0004-6361/201629272}

\bibitem[{{Gaia Collaboration} {et~al.}(2018){Gaia Collaboration}, {Helmi},
  {van Leeuwen}, {McMillan}, {Massari}, {Antoja}, {Robin}, {Lindegren},
  {Bastian}, {Arenou}, {Babusiaux}, {Biermann}, {Breddels}, {Hobbs}, {Jordi},
  {Pancino}, {Reyl{\'e}}, {Veljanoski}, {Brown}, {Vallenari}, {Prusti}, {de
  Bruijne}, {Bailer-Jones}, {Evans}, {Eyer}, {Jansen}, {Klioner}, {Lammers},
  {Luri}, {Mignard}, {Panem}, {Pourbaix}, {Randich}, {Sartoretti}, {Siddiqui},
  {Soubiran}, {Walton}, {Cropper}, {Drimmel}, {Katz}, {Lattanzi}, {Bakker},
  {Cacciari}, {Casta{\~n}eda}, {Chaoul}, {Cheek}, {De Angeli}, {Fabricius},
  {Guerra}, {Holl}, {Masana}, {Messineo}, {Mowlavi}, {Nienartowicz}, {Panuzzo},
  {Portell}, {Riello}, {Seabroke}, {Tanga}, {Th{\'e}venin}, {Gracia-Abril},
  {Comoretto}, {Garcia-Reinaldos}, {Teyssier}, {Altmann}, {Andrae}, {Audard},
  {Bellas-Velidis}, {Benson}, {Berthier}, {Blomme}, {Burgess}, {Busso},
  {Carry}, {Cellino}, {Clementini}, {Clotet}, {Creevey}, {Davidson}, {De
  Ridder}, {Delchambre}, {Dell'Oro}, {Ducourant},
  {Fern{\'a}ndez-Hern{\'a}ndez}, {Fouesneau}, {Fr{\'e}mat}, {Galluccio},
  {Garc{\'\i}a-Torres}, {Gonz{\'a}lez-N{\'u}{\~n}ez}, {Gonz{\'a}lez-Vidal},
  {Gosset}, {Guy}, {Halbwachs}, {Hambly}, {Harrison}, {Hern{\'a}ndez},
  {Hestroffer}, {Hodgkin}, {Hutton}, {Jasniewicz}, {Jean-Antoine-Piccolo},
  {Jordan}, {Korn}, {Krone-Martins}, {Lanzafame}, {Lebzelter}, {L{\"o}ffler},
  {Manteiga}, {Marrese}, {Mart{\'\i}n-Fleitas}, {Moitinho}, {Mora}, {Muinonen},
  {Osinde}, {Pauwels}, {Petit}, {Recio-Blanco}, {Richards}, {Rimoldini},
  {Sarro}, {Siopis}, {Smith}, {Sozzetti}, {S{\"u}veges}, {Torra}, {van Reeven},
  {Abbas}, {Abreu Aramburu}, {Accart}, {Aerts}, {Altavilla}, {{\'A}lvarez},
  {Alvarez}, {Alves}, {Anderson}, {Andrei}, {Anglada Varela}, {Antiche},
  {Arcay}, {Astraatmadja}, {Bach}, {Baker}, {Balaguer-N{\'u}{\~n}ez}, {Balm},
  {Barache}, {Barata}, {Barbato}, {Barblan}, {Barklem}, {Barrado}, {Barros},
  {Barstow}, {Bartholom{\'e} Mu{\~n}oz}, {Bassilana}, {Becciani}, {Bellazzini},
  {Berihuete}, {Bertone}, {Bianchi}, {Bienaym{\'e}}, {Blanco-Cuaresma}, {Boch},
  {Boeche}, {Bombrun}, {Borrachero}, {Bossini}, {Bouquillon}, {Bourda},
  {Bragaglia}, {Bramante}, {Bressan}, {Brouillet}, {Br{\"u}semeister},
  {Brugaletta}, {Bucciarelli}, {Burlacu}, {Busonero}, {Butkevich}, {Buzzi},
  {Caffau}, {Cancelliere}, {Cannizzaro}, {Cantat-Gaudin}, {Carballo},
  {Carlucci}, {Carrasco}, {Casamiquela}, {Castellani}, {Castro-Ginard},
  {Charlot}, {Chemin}, {Chiavassa}, {Cocozza}, {Costigan}, {Cowell}, {Crifo},
  {Crosta}, {Crowley}, {Cuypers}, {Dafonte}, {Damerdji}, {Dapergolas}, {David},
  {David}, {de Laverny}, {De Luise}, {De March}, {de Martino}, {de Souza}, {de
  Torres}, {Debosscher}, {del Pozo}, {Delbo}, {Delgado}, {Delgado}, {Di
  Matteo}, {Diakite}, {Diener}, {Distefano}, {Dolding}, {Drazinos},
  {Dur{\'a}n}, {Edvardsson}, {Enke}, {Eriksson}, {Esquej}, {Eynard Bontemps},
  {Fabre}, {Fabrizio}, {Faigler}, {Falc{\~a}o}, {Farr{\`a}s Casas}, {Federici},
  {Fedorets}, {Fernique}, {Figueras}, {Filippi}, {Findeisen}, {Fonti},
  {Fraile}, {Fraser}, {Fr{\'e}zouls}, {Gai}, {Galleti}, {Garabato},
  {Garc{\'\i}a-Sedano}, {Garofalo}, {Garralda}, {Gavel}, {Gavras}, {Gerssen},
  {Geyer}, {Giacobbe}, {Gilmore}, {Girona}, {Giuffrida}, {Glass}, {Gomes},
  {Granvik}, {Gueguen}, {Guerrier}, {Guiraud}, {Guti{\'e}rrez-S{\'a}nchez},
  {Hofmann}, {Holland}, {Huckle}, {Hypki}, {Icardi}, {Jan{\ss}en}, {Jevardat de
  Fombelle}, {Jonker}, {Juh{\'a}sz}, {Julbe}, {Karampelas}, {Kewley}, {Klar},
  {Kochoska}, {Kohley}, {Kolenberg}, {Kontizas}, {Kontizas}, {Koposov},
  {Kordopatis}, {Kostrzewa-Rutkowska}, {Koubsky}, {Lambert}, {Lanza}, {Lasne},
  {Lavigne}, {Le Fustec}, {Le Poncin-Lafitte}, {Lebreton}, {Leccia}, {Leclerc},
  {Lecoeur-Taibi}, {Lenhardt}, {Leroux}, {Liao}, {Licata}, {Lindstr{\o}m},
  {Lister}, {Livanou}, {Lobel}, {L{\'o}pez}, {Managau}, {Mann}, {Mantelet},
  {Marchal}, {Marchant}, {Marconi}, {Marinoni}, {Marschalk{\'o}}, {Marshall},
  {Martino}, {Marton}, {Mary}, {Matijevi{\v{c}}}, {Mazeh}, {Messina},
  {Michalik}, {Millar}, {Molina}, {Molinaro}, {Moln{\'a}r}, {Montegriffo},
  {Mor}, {Morbidelli}, {Morel}, {Morris}, {Mulone}, {Muraveva}, {Musella},
  {Nelemans}, {Nicastro}, {Noval}, {O'Mullane}, {Ord{\'e}novic},
  {Ord{\'o}{\~n}ez-Blanco}, {Osborne}, {Pagani}, {Pagano}, {Pailler},
  {Palacin}, {Palaversa}, {Panahi}, {Pawlak}, {Piersimoni}, {Pineau}, {Plachy},
  {Plum}, {Poggio}, {Poujoulet}, {Pr{\v{s}}a}, {Pulone}, {Racero}, {Ragaini},
  {Rambaux}, {Ramos-Lerate}, {Regibo}, {Riclet}, {Ripepi}, {Riva}, {Rivard},
  {Rixon}, {Roegiers}, {Roelens}, {Romero-G{\'o}mez}, {Rowell}, {Royer},
  {Ruiz-Dern}, {Sadowski}, {Sagrist{\`a} Sell{\'e}s}, {Sahlmann}, {Salgado},
  {Salguero}, {Sanna}, {Santana-Ros}, {Sarasso}, {Savietto}, {Schultheis},
  {Sciacca}, {Segol}, {Segovia}, {S{\'e}gransan}, {Shih}, {Siltala}, {Silva},
  {Smart}, {Smith}, {Solano}, {Solitro}, {Sordo}, {Soria Nieto}, {Souchay},
  {Spagna}, {Spoto}, {Stampa}, {Steele}, {Steidelm{\"u}ller}, {Stephenson},
  {Stoev}, {Suess}, {Surdej}, {Szabados}, {Szegedi-Elek}, {Tapiador}, {Taris},
  {Tauran}, {Taylor}, {Teixeira}, {Terrett}, {Teyssandier}, {Thuillot},
  {Titarenko}, {Torra Clotet}, {Turon}, {Ulla}, {Utrilla}, {Uzzi}, {Vaillant},
  {Valentini}, {Valette}, {van Elteren}, {Van Hemelryck}, {van Leeuwen},
  {Vaschetto}, {Vecchiato}, {Viala}, {Vicente}, {Vogt}, {von Essen}, {Voss},
  {Votruba}, {Voutsinas}, {Walmsley}, {Weiler}, {Wertz}, {Wevems},
  {Wyrzykowski}, {Yoldas}, {{\v{Z}}erjal}, {Ziaeepour}, {Zorec}, {Zschocke},
  {Zucker}, {Zurbach}, \& {Zwitter}}]{gaia18}
{Gaia Collaboration}, {Helmi}, A., {van Leeuwen}, F., {et~al.} 2018, \aap, 616,
  A12, \dodoi{10.1051/0004-6361/201832698}

\bibitem[{{Gaia Collaboration} {et~al.}(2021){Gaia Collaboration}, {Brown, A.
  G. A.}, {Vallenari, A.}, {Prusti, T.}, {de Bruijne, J. H. J.}, {Babusiaux,
  C.}, {Biermann, M.}, {Creevey, O. L.}, {Evans, D. W.}, {Eyer, L.}, {Hutton,
  A.}, {Jansen, F.}, {Jordi, C.}, {Klioner, S. A.}, {Lammers, U.}, {Lindegren,
  L.}, {Luri, X.}, {Mignard, F.}, {Panem, C.}, {Pourbaix, D.}, {Randich, S.},
  {Sartoretti, P.}, {Soubiran, C.}, {Walton, N. A.}, {Arenou, F.},
  {Bailer-Jones, C. A. L.}, {Bastian, U.}, {Cropper, M.}, {Drimmel, R.}, {Katz,
  D.}, {Lattanzi, M. G.}, {van Leeuwen, F.}, {Bakker, J.}, {Cacciari, C.},
  {Casta\~neda, J.}, {De Angeli, F.}, {Ducourant, C.}, {Fabricius, C.},
  {Fouesneau, M.}, {Fr\'emat, Y.}, {Guerra, R.}, {Guerrier, A.}, {Guiraud, J.},
  {Jean-Antoine Piccolo, A.}, {Masana, E.}, {Messineo, R.}, {Mowlavi, N.},
  {Nicolas, C.}, {Nienartowicz, K.}, {Pailler, F.}, {Panuzzo, P.}, {Riclet,
  F.}, {Roux, W.}, {Seabroke, G. M.}, {Sordo, R.}, {Tanga, P.}, {Th\'evenin,
  F.}, {Gracia-Abril, G.}, {Portell, J.}, {Teyssier, D.}, {Altmann, M.},
  {Andrae, R.}, {Bellas-Velidis, I.}, {Benson, K.}, {Berthier, J.}, {Blomme,
  R.}, {Brugaletta, E.}, {Burgess, P. W.}, {Busso, G.}, {Carry, B.}, {Cellino,
  A.}, {Cheek, N.}, {Clementini, G.}, {Damerdji, Y.}, {Davidson, M.},
  {Delchambre, L.}, {Dell\'{}Oro, A.}, {Fern\'andez-Hern\'andez, J.},
  {Galluccio, L.}, {Garc\'{\i}a-Lario, P.}, {Garcia-Reinaldos, M.},
  {Gonz\'alez-N\'u\~nez, J.}, {Gosset, E.}, {Haigron, R.}, {Halbwachs, J.-L.},
  {Hambly, N. C.}, {Harrison, D. L.}, {Hatzidimitriou, D.}, {Heiter, U.},
  {Hern\'andez, J.}, {Hestroffer, D.}, {Hodgkin, S. T.}, {Holl, B.},
  {Jan\ss{}en, K.}, {Jevardat de Fombelle, G.}, {Jordan, S.}, {Krone-Martins,
  A.}, {Lanzafame, A. C.}, {L\"offler, W.}, {Lorca, A.}, {Manteiga, M.},
  {Marchal, O.}, {Marrese, P. M.}, {Moitinho, A.}, {Mora, A.}, {Muinonen, K.},
  {Osborne, P.}, {Pancino, E.}, {Pauwels, T.}, {Petit, J.-M.}, {Recio-Blanco,
  A.}, {Richards, P. J.}, {Riello, M.}, {Rimoldini, L.}, {Robin, A. C.},
  {Roegiers, T.}, {Rybizki, J.}, {Sarro, L. M.}, {Siopis, C.}, {Smith, M.},
  {Sozzetti, A.}, {Ulla, A.}, {Utrilla, E.}, {van Leeuwen, M.}, {van Reeven,
  W.}, {Abbas, U.}, {Abreu Aramburu, A.}, {Accart, S.}, {Aerts, C.}, {Aguado,
  J. J.}, {Ajaj, M.}, {Altavilla, G.}, {\'Alvarez, M. A.}, {\'Alvarez
  Cid-Fuentes, J.}, {Alves, J.}, {Anderson, R. I.}, {Anglada Varela, E.},
  {Antoja, T.}, {Audard, M.}, {Baines, D.}, {Baker, S. G.},
  {Balaguer-N\'u\~nez, L.}, {Balbinot, E.}, {Balog, Z.}, {Barache, C.},
  {Barbato, D.}, {Barros, M.}, {Barstow, M. A.}, {Bartolom\'e, S.}, {Bassilana,
  J.-L.}, {Bauchet, N.}, {Baudesson-Stella, A.}, {Becciani, U.}, {Bellazzini,
  M.}, {Bernet, M.}, {Bertone, S.}, {Bianchi, L.}, {Blanco-Cuaresma, S.},
  {Boch, T.}, {Bombrun, A.}, {Bossini, D.}, {Bouquillon, S.}, {Bragaglia, A.},
  {Bramante, L.}, {Breedt, E.}, {Bressan, A.}, {Brouillet, N.}, {Bucciarelli,
  B.}, {Burlacu, A.}, {Busonero, D.}, {Butkevich, A. G.}, {Buzzi, R.}, {Caffau,
  E.}, {Cancelliere, R.}, {C\'anovas, H.}, {Cantat-Gaudin, T.}, {Carballo, R.},
  {Carlucci, T.}, {Carnerero, M. I}, {Carrasco, J. M.}, {Casamiquela, L.},
  {Castellani, M.}, {Castro-Ginard, A.}, {Castro Sampol, P.}, {Chaoul, L.},
  {Charlot, P.}, {Chemin, L.}, {Chiavassa, A.}, {Cioni, M.-R. L.}, {Comoretto,
  G.}, {Cooper, W. J.}, {Cornez, T.}, {Cowell, S.}, {Crifo, F.}, {Crosta, M.},
  {Crowley, C.}, {Dafonte, C.}, {Dapergolas, A.}, {David, M.}, {David, P.}, {de
  Laverny, P.}, {De Luise, F.}, {De March, R.}, {De Ridder, J.}, {de Souza,
  R.}, {de Teodoro, P.}, {de Torres, A.}, {del Peloso, E. F.}, {del Pozo, E.},
  {Delbo, M.}, {Delgado, A.}, {Delgado, H. E.}, {Delisle, J.-B.}, {Di Matteo,
  P.}, {Diakite, S.}, {Diener, C.}, {Distefano, E.}, {Dolding, C.}, {Eappachen,
  D.}, {Edvardsson, B.}, {Enke, H.}, {Esquej, P.}, {Fabre, C.}, {Fabrizio, M.},
  {Faigler, S.}, {Fedorets, G.}, {Fernique, P.}, {Fienga, A.}, {Figueras, F.},
  {Fouron, C.}, {Fragkoudi, F.}, {Fraile, E.}, {Franke, F.}, {Gai, M.},
  {Garabato, D.}, {Garcia-Gutierrez, A.}, {Garc\'{\i}a-Torres, M.}, {Garofalo,
  A.}, {Gavras, P.}, {Gerlach, E.}, {Geyer, R.}, {Giacobbe, P.}, {Gilmore, G.},
  {Girona, S.}, {Giuffrida, G.}, {Gomel, R.}, {Gomez, A.},
  {Gonzalez-Santamaria, I.}, {Gonz\'alez-Vidal, J. J.}, {Granvik, M.},
  {Guti\'errez-S\'anchez, R.}, {Guy, L. P.}, {Hauser, M.}, {Haywood, M.},
  {Helmi, A.}, {Hidalgo, S. L.}, {Hilger, T.}, {Hladczuk, N.}, {Hobbs, D.},
  {Holland, G.}, {Huckle, H. E.}, {Jasniewicz, G.}, {Jonker, P. G.}, {Juaristi
  Campillo, J.}, {Julbe, F.}, {Karbevska, L.}, {Kervella, P.}, {Khanna, S.},
  {Kochoska, A.}, {Kontizas, M.}, {Kordopatis, G.}, {Korn, A. J.},
  {Kostrzewa-Rutkowska, Z.}, {Kruszy\'{}nska, K.}, {Lambert, S.}, {Lanza, A.
  F.}, {Lasne, Y.}, {Le Campion, J.-F.}, {Le Fustec, Y.}, {Lebreton, Y.},
  {Lebzelter, T.}, {Leccia, S.}, {Leclerc, N.}, {Lecoeur-Taibi, I.}, {Liao,
  S.}, {Licata, E.}, {Lindstr\o{}m, E. P.}, {Lister, T. A.}, {Livanou, E.},
  {Lobel, A.}, {Madrero Pardo, P.}, {Managau, S.}, {Mann, R. G.}, {Marchant, J.
  M.}, {Marconi, M.}, {Marcos Santos, M. M. S.}, {Marinoni, S.}, {Marocco, F.},
  {Marshall, D. J.}, {Martin Polo, L.}, {Mart\'{\i}n-Fleitas, J. M.}, {Masip,
  A.}, {Massari, D.}, {Mastrobuono-Battisti, A.}, {Mazeh, T.}, {McMillan, P.
  J.}, {Messina, S.}, {Michalik, D.}, {Millar, N. R.}, {Mints, A.}, {Molina,
  D.}, {Molinaro, R.}, {Moln\'ar, L.}, {Montegriffo, P.}, {Mor, R.},
  {Morbidelli, R.}, {Morel, T.}, {Morris, D.}, {Mulone, A. F.}, {Munoz, D.},
  {Muraveva, T.}, {Murphy, C. P.}, {Musella, I.}, {Noval, L.}, {Ord\'enovic,
  C.}, {Orr\`u, G.}, {Osinde, J.}, {Pagani, C.}, {Pagano, I.}, {Palaversa, L.},
  {Palicio, P. A.}, {Panahi, A.}, {Pawlak, M.}, {Pe\~nalosa Esteller, X.},
  {Penttil\"a, A.}, {Piersimoni, A. M.}, {Pineau, F.-X.}, {Plachy, E.}, {Plum,
  G.}, {Poggio, E.}, {Poretti, E.}, {Poujoulet, E.}, {Prsa, A.}, {Pulone, L.},
  {Racero, E.}, {Ragaini, S.}, {Rainer, M.}, {Raiteri, C. M.}, {Rambaux, N.},
  {Ramos, P.}, {Ramos-Lerate, M.}, {Re Fiorentin, P.}, {Regibo, S.}, {Reyl\'e,
  C.}, {Ripepi, V.}, {Riva, A.}, {Rixon, G.}, {Robichon, N.}, {Robin, C.},
  {Roelens, M.}, {Rohrbasser, L.}, {Romero-G\'omez, M.}, {Rowell, N.}, {Royer,
  F.}, {Rybicki, K. A.}, {Sadowski, G.}, {Sagrist\`a Sell\'es, A.}, {Sahlmann,
  J.}, {Salgado, J.}, {Salguero, E.}, {Samaras, N.}, {Sanchez Gimenez, V.},
  {Sanna, N.}, {Santove\~na, R.}, {Sarasso, M.}, {Schultheis, M.}, {Sciacca,
  E.}, {Segol, M.}, {Segovia, J. C.}, {S\'egransan, D.}, {Semeux, D.}, {Shahaf,
  S.}, {Siddiqui, H. I.}, {Siebert, A.}, {Siltala, L.}, {Slezak, E.}, {Smart,
  R. L.}, {Solano, E.}, {Solitro, F.}, {Souami, D.}, {Souchay, J.}, {Spagna,
  A.}, {Spoto, F.}, {Steele, I. A.}, {Steidelm\"uller, H.}, {Stephenson, C.
  A.}, {S\"uveges, M.}, {Szabados, L.}, {Szegedi-Elek, E.}, {Taris, F.},
  {Tauran, G.}, {Taylor, M. B.}, {Teixeira, R.}, {Thuillot, W.}, {Tonello, N.},
  {Torra, F.}, {Torra, J.}, {Turon, C.}, {Unger, N.}, {Vaillant, M.}, {van
  Dillen, E.}, {Vanel, O.}, {Vecchiato, A.}, {Viala, Y.}, {Vicente, D.},
  {Voutsinas, S.}, {Weiler, M.}, {Wevers, T.}, {Wyrzykowski, L.}, {Yoldas, A.},
  {Yvard, P.}, {Zhao, H.}, {Zorec, J.}, {Zucker, S.}, {Zurbach, C.}, \&
  {Zwitter, T.}}]{gaia21}
{Gaia Collaboration}, {Brown, A. G. A.}, {Vallenari, A.}, {et~al.} 2021, A\&A,
  649, A1, \dodoi{10.1051/0004-6361/202039657}

\bibitem[{{Garc{\'\i}a P{\'e}rez} {et~al.}(2016){Garc{\'\i}a P{\'e}rez},
  {Allende Prieto}, {Holtzman}, {Shetrone}, {M{\'e}sz{\'a}ros}, {Bizyaev},
  {Carrera}, {Cunha}, {Garc{\'\i}a-Hern{\'a}ndez}, {Johnson}, {Majewski},
  {Nidever}, {Schiavon}, {Shane}, {Smith}, {Sobeck}, {Troup}, {Zamora},
  {Weinberg}, {Bovy}, {Eisenstein}, {Feuillet}, {Frinchaboy}, {Hayden},
  {Hearty}, {Nguyen}, {O'Connell}, {Pinsonneault}, {Wilson}, \&
  {Zasowski}}]{gar16}
{Garc{\'\i}a P{\'e}rez}, A.~E., {Allende Prieto}, C., {Holtzman}, J.~A.,
  {et~al.} 2016, \aj, 151, 144, \dodoi{10.3847/0004-6256/151/6/144}

\bibitem[{{Gratton} {et~al.}(2004){Gratton}, {Sneden}, \& {Carretta}}]{gra04}
{Gratton}, R., {Sneden}, C., \& {Carretta}, E. 2004, \araa, 42, 385,
  \dodoi{10.1146/annurev.astro.42.053102.133945}

\bibitem[{{Gratton}(1983)}]{gra83}
{Gratton}, R.~G. 1983, \aap, 123, 289

\bibitem[{{Gratton} \& {Ortolani}(1986)}]{gra86}
{Gratton}, R.~G., \& {Ortolani}, S. 1986, \aap, 169, 201

\bibitem[{{Gratton} \& {Sneden}(1988)}]{gra88}
{Gratton}, R.~G., \& {Sneden}, C. 1988, \aap, 204, 193

\bibitem[{{Gratton} {et~al.}(2001){Gratton}, {Bonifacio}, {Bragaglia},
  {Carretta}, {Castellani}, {Centurion}, {Chieffi}, {Claudi}, {Clementini},
  {D'Antona}, {Desidera}, {Fran{\c{c}}ois}, {Grundahl}, {Lucatello}, {Molaro},
  {Pasquini}, {Sneden}, {Spite}, \& {Straniero}}]{gra01}
{Gratton}, R.~G., {Bonifacio}, P., {Bragaglia}, A., {et~al.} 2001, \aap, 369,
  87, \dodoi{10.1051/0004-6361:20010144}

\bibitem[{{Graur} {et~al.}(2014){Graur}, {Rodney}, {Maoz}, {Riess}, {Jha},
  {Postman}, {Dahlen}, {Holoien}, {McCully}, {Patel}, {Strolger},
  {Ben{\'\i}tez}, {Coe}, {Jouvel}, {Medezinski}, {Molino}, {Nonino}, {Bradley},
  {Koekemoer}, {Balestra}, {Cenko}, {Clubb}, {Dickinson}, {Filippenko},
  {Frederiksen}, {Garnavich}, {Hjorth}, {Jones}, {Leibundgut}, {Matheson},
  {Mobasher}, {Rosati}, {Silverman}, {U}, {Jedruszczuk}, {Li}, {Lin},
  {Mirmelstein}, {Neustadt}, {Ovadia}, \& {Rogers}}]{gra14}
{Graur}, O., {Rodney}, S.~A., {Maoz}, D., {et~al.} 2014, \apj, 783, 28,
  \dodoi{10.1088/0004-637X/783/1/28}

\bibitem[{{Gronow} {et~al.}(2021{\natexlab{a}}){Gronow}, {Collins}, {Sim}, \&
  {R{\"o}pke}}]{gro21a}
{Gronow}, S., {Collins}, C.~E., {Sim}, S.~A., \& {R{\"o}pke}, F.~K.
  2021{\natexlab{a}}, \aap, 649, A155, \dodoi{10.1051/0004-6361/202039954}

\bibitem[{{Gronow} {et~al.}(2021{\natexlab{b}}){Gronow}, {C{\^o}t{\'e}},
  {Lach}, {Seitenzahl}, {Collins}, {Sim}, \& {R{\"o}pke}}]{gro21b}
{Gronow}, S., {C{\^o}t{\'e}}, B., {Lach}, F., {et~al.} 2021{\natexlab{b}},
  \aap, 656, A94, \dodoi{10.1051/0004-6361/202140881}

\bibitem[{{Gunn} {et~al.}(2006){Gunn}, {Siegmund}, {Mannery}, {Owen}, {Hull},
  {Leger}, {Carey}, {Knapp}, {York}, {Boroski}, {Kent}, {Lupton}, {Rockosi},
  {Evans}, {Waddell}, {Anderson}, {Annis}, {Barentine}, {Bartoszek}, {Bastian},
  {Bracker}, {Brewington}, {Briegel}, {Brinkmann}, {Brown}, {Carr},
  {Czarapata}, {Drennan}, {Dombeck}, {Federwitz}, {Gillespie}, {Gonzales},
  {Hansen}, {Harvanek}, {Hayes}, {Jordan}, {Kinney}, {Klaene}, {Kleinman},
  {Kron}, {Kresinski}, {Lee}, {Limmongkol}, {Lindenmeyer}, {Long}, {Loomis},
  {McGehee}, {Mantsch}, {Neilsen}, {Neswold}, {Newman}, {Nitta}, {Peoples},
  {Pier}, {Prieto}, {Prosapio}, {Rivetta}, {Schneider}, {Snedden}, \&
  {Wang}}]{gun06}
{Gunn}, J.~E., {Siegmund}, W.~A., {Mannery}, E.~J., {et~al.} 2006, \aj, 131,
  2332, \dodoi{10.1086/500975}

\bibitem[{{Gustafsson} {et~al.}(2008){Gustafsson}, {Edvardsson}, {Eriksson},
  {J{\o}rgensen}, {Nordlund}, \& {Plez}}]{gus08}
{Gustafsson}, B., {Edvardsson}, B., {Eriksson}, K., {et~al.} 2008, \aap, 486,
  951, \dodoi{10.1051/0004-6361:200809724}

\bibitem[{{Hansen} {et~al.}(2018){Hansen}, {Holmbeck}, {Beers}, {Placco},
  {Roederer}, {Frebel}, {Sakari}, {Simon}, \& {Thompson}}]{han18}
{Hansen}, T.~T., {Holmbeck}, E.~M., {Beers}, T.~C., {et~al.} 2018, \apj, 858,
  92, \dodoi{10.3847/1538-4357/aabacc}

\bibitem[{{Harris} {et~al.}(2020){Harris}, {Millman}, {van der Walt},
  {Gommers}, {Virtanen}, {Cournapeau}, {Wieser}, {Taylor}, {Berg}, {Smith},
  {Kern}, {Picus}, {Hoyer}, {van Kerkwijk}, {Brett}, {Haldane}, {del R{\'\i}o},
  {Wiebe}, {Peterson}, {G{\'e}rard-Marchant}, {Sheppard}, {Reddy}, {Weckesser},
  {Abbasi}, {Gohlke}, \& {Oliphant}}]{har20}
{Harris}, C.~R., {Millman}, K.~J., {van der Walt}, S.~J., {et~al.} 2020, \nat,
  585, 357, \dodoi{10.1038/s41586-020-2649-2}

\bibitem[{{Hayes} {et~al.}(2018){Hayes}, {Majewski}, {Shetrone},
  {Fern{\'a}ndez-Alvar}, {Allende Prieto}, {Schuster}, {Carigi}, {Cunha},
  {Smith}, {Sobeck}, {Almeida}, {Beers}, {Carrera}, {Fern{\'a}ndez-Trincado},
  {Garc{\'\i}a-Hern{\'a}ndez}, {Geisler}, {Lane}, {Lucatello}, {Matthews},
  {Minniti}, {Nitschelm}, {Tang}, {Tissera}, \& {Zamora}}]{hay18}
{Hayes}, C.~R., {Majewski}, S.~R., {Shetrone}, M., {et~al.} 2018, \apj, 852,
  49, \dodoi{10.3847/1538-4357/aa9cec}

\bibitem[{{Heger} \& {Woosley}(2002)}]{heg02}
{Heger}, A., \& {Woosley}, S.~E. 2002, \apj, 567, 532, \dodoi{10.1086/338487}

\bibitem[{{Holtzman} {et~al.}(2010){Holtzman}, {Harrison}, \&
  {Coughlin}}]{hol10}
{Holtzman}, J.~A., {Harrison}, T.~E., \& {Coughlin}, J.~L. 2010, Advances in
  Astronomy, 2010, 193086, \dodoi{10.1155/2010/193086}

\bibitem[{{Holtzman} {et~al.}(2015){Holtzman}, {Shetrone}, {Johnson}, {Allende
  Prieto}, {Anders}, {Andrews}, {Beers}, {Bizyaev}, {Blanton}, {Bovy},
  {Carrera}, {Chojnowski}, {Cunha}, {Eisenstein}, {Feuillet}, {Frinchaboy},
  {Galbraith-Frew}, {Garc{\'\i}a P{\'e}rez}, {Garc{\'\i}a-Hern{\'a}ndez},
  {Hasselquist}, {Hayden}, {Hearty}, {Ivans}, {Majewski}, {Martell},
  {Meszaros}, {Muna}, {Nidever}, {Nguyen}, {O'Connell}, {Pan}, {Pinsonneault},
  {Robin}, {Schiavon}, {Shane}, {Sobeck}, {Smith}, {Troup}, {Weinberg},
  {Wilson}, {Wood-Vasey}, {Zamora}, \& {Zasowski}}]{hol15}
{Holtzman}, J.~A., {Shetrone}, M., {Johnson}, J.~A., {et~al.} 2015, \aj, 150,
  148, \dodoi{10.1088/0004-6256/150/5/148}

\bibitem[{{Hubeny} \& {Lanz}(2011)}]{hub11}
{Hubeny}, I., \& {Lanz}, T. 2011, {Synspec: General Spectrum Synthesis
  Program}, Astrophysics Source Code Library, record ascl:1109.022.
\newblock \doeprint{1109.022}

\bibitem[{Hunter(2007)}]{hun07}
Hunter, J.~D. 2007, Computing in Science \& Engineering, 9, 90,
  \dodoi{10.1109/MCSE.2007.55}

\bibitem[{{J{\"o}nsson} {et~al.}(2020){J{\"o}nsson}, {Holtzman}, {Allende
  Prieto}, {Cunha}, {Garc{\'\i}a-Hern{\'a}ndez}, {Hasselquist}, {Masseron},
  {Osorio}, {Shetrone}, {Smith}, {Stringfellow}, {Bizyaev}, {Edvardsson},
  {Majewski}, {M{\'e}sz{\'a}ros}, {Souto}, {Zamora}, {Beaton}, {Bovy}, {Donor},
  {Pinsonneault}, {Poovelil}, \& {Sobeck}}]{jon20}
{J{\"o}nsson}, H., {Holtzman}, J.~A., {Allende Prieto}, C., {et~al.} 2020, \aj,
  160, 120, \dodoi{10.3847/1538-3881/aba592}

\bibitem[{{Kirby} {et~al.}(2011{\natexlab{a}}){Kirby}, {Cohen}, {Smith},
  {Majewski}, {Sohn}, \& {Guhathakurta}}]{kir11a}
{Kirby}, E.~N., {Cohen}, J.~G., {Smith}, G.~H., {et~al.} 2011{\natexlab{a}},
  \apj, 727, 79, \dodoi{10.1088/0004-637X/727/2/79}

\bibitem[{{Kirby} {et~al.}(2009){Kirby}, {Guhathakurta}, {Bolte}, {Sneden}, \&
  {Geha}}]{kir09}
{Kirby}, E.~N., {Guhathakurta}, P., {Bolte}, M., {Sneden}, C., \& {Geha}, M.~C.
  2009, \apj, 705, 328, \dodoi{10.1088/0004-637X/705/1/328}

\bibitem[{{Kirby} {et~al.}(2011{\natexlab{b}}){Kirby}, {Lanfranchi}, {Simon},
  {Cohen}, \& {Guhathakurta}}]{kir11b}
{Kirby}, E.~N., {Lanfranchi}, G.~A., {Simon}, J.~D., {Cohen}, J.~G., \&
  {Guhathakurta}, P. 2011{\natexlab{b}}, \apj, 727, 78,
  \dodoi{10.1088/0004-637X/727/2/78}

\bibitem[{{Kirby} {et~al.}(2010){Kirby}, {Guhathakurta}, {Simon}, {Geha},
  {Rockosi}, {Sneden}, {Cohen}, {Sohn}, {Majewski}, \& {Siegel}}]{kir10}
{Kirby}, E.~N., {Guhathakurta}, P., {Simon}, J.~D., {et~al.} 2010, \apjs, 191,
  352, \dodoi{10.1088/0067-0049/191/2/352}

\bibitem[{{Lacchin} {et~al.}(2021){Lacchin}, {Calura}, \& {Vesperini}}]{lal21}
{Lacchin}, E., {Calura}, F., \& {Vesperini}, E. 2021, \mnras, 506, 5951,
  \dodoi{10.1093/mnras/stab2061}

\bibitem[{{Lanfranchi} \& {Matteucci}(2003)}]{lan03}
{Lanfranchi}, G.~A., \& {Matteucci}, F. 2003, \mnras, 345, 71,
  \dodoi{10.1046/j.1365-8711.2003.06919.x}

\bibitem[{{Lanfranchi} \& {Matteucci}(2004)}]{lan04}
---. 2004, \mnras, 351, 1338, \dodoi{10.1111/j.1365-2966.2004.07877.x}

\bibitem[{{Lanfranchi} \& {Matteucci}(2007)}]{lan07}
---. 2007, \aap, 468, 927, \dodoi{10.1051/0004-6361:20066576}

\bibitem[{{Lanfranchi} {et~al.}(2006){Lanfranchi}, {Matteucci}, \&
  {Cescutti}}]{lan06}
{Lanfranchi}, G.~A., {Matteucci}, F., \& {Cescutti}, G. 2006, \aap, 453, 67,
  \dodoi{10.1051/0004-6361:20054627}

\bibitem[{{Lanfranchi} {et~al.}(2008){Lanfranchi}, {Matteucci}, \&
  {Cescutti}}]{lan08}
---. 2008, \aap, 481, 635, \dodoi{10.1051/0004-6361:20078696}

\bibitem[{{Leung} \& {Nomoto}(2018)}]{leu18}
{Leung}, S.-C., \& {Nomoto}, K. 2018, \apj, 861, 143,
  \dodoi{10.3847/1538-4357/aac2df}

\bibitem[{{Leung} \& {Nomoto}(2020{\natexlab{a}})}]{leu20a}
---. 2020{\natexlab{a}}, \apj, 888, 80, \dodoi{10.3847/1538-4357/ab5c1f}

\bibitem[{{Leung} \& {Nomoto}(2020{\natexlab{b}})}]{leu20b}
---. 2020{\natexlab{b}}, \apj, 900, 54, \dodoi{10.3847/1538-4357/aba1e3}

\bibitem[{{Lindegren} {et~al.}(2018){Lindegren}, {Hern{\'a}ndez}, {Bombrun},
  {Klioner}, {Bastian}, {Ramos-Lerate}, {de Torres}, {Steidelm{\"u}ller},
  {Stephenson}, {Hobbs}, {Lammers}, {Biermann}, {Geyer}, {Hilger}, {Michalik},
  {Stampa}, {McMillan}, {Casta{\~n}eda}, {Clotet}, {Comoretto}, {Davidson},
  {Fabricius}, {Gracia}, {Hambly}, {Hutton}, {Mora}, {Portell}, {van Leeuwen},
  {Abbas}, {Abreu}, {Altmann}, {Andrei}, {Anglada}, {Balaguer-N{\'u}{\~n}ez},
  {Barache}, {Becciani}, {Bertone}, {Bianchi}, {Bouquillon}, {Bourda},
  {Br{\"u}semeister}, {Bucciarelli}, {Busonero}, {Buzzi}, {Cancelliere},
  {Carlucci}, {Charlot}, {Cheek}, {Crosta}, {Crowley}, {de Bruijne}, {de
  Felice}, {Drimmel}, {Esquej}, {Fienga}, {Fraile}, {Gai}, {Garralda},
  {Gonz{\'a}lez-Vidal}, {Guerra}, {Hauser}, {Hofmann}, {Holl}, {Jordan},
  {Lattanzi}, {Lenhardt}, {Liao}, {Licata}, {Lister}, {L{\"o}ffler},
  {Marchant}, {Martin-Fleitas}, {Messineo}, {Mignard}, {Morbidelli}, {Poggio},
  {Riva}, {Rowell}, {Salguero}, {Sarasso}, {Sciacca}, {Siddiqui}, {Smart},
  {Spagna}, {Steele}, {Taris}, {Torra}, {van Elteren}, {van Reeven}, \&
  {Vecchiato}}]{lin18}
{Lindegren}, L., {Hern{\'a}ndez}, J., {Bombrun}, A., {et~al.} 2018, \aap, 616,
  A2, \dodoi{10.1051/0004-6361/201832727}

\bibitem[{{Lindegren} {et~al.}(2021{\natexlab{a}}){Lindegren}, {Bastian},
  {Biermann}, {Bombrun}, {de Torres}, {Gerlach}, {Geyer}, {Hern{\'a}ndez},
  {Hilger}, {Hobbs}, {Klioner}, {Lammers}, {McMillan}, {Ramos-Lerate},
  {Steidelm{\"u}ller}, {Stephenson}, \& {van Leeuwen}}]{lin21a}
{Lindegren}, L., {Bastian}, U., {Biermann}, M., {et~al.} 2021{\natexlab{a}},
  \aap, 649, A4, \dodoi{10.1051/0004-6361/202039653}

\bibitem[{{Lindegren} {et~al.}(2021{\natexlab{b}}){Lindegren}, {Klioner},
  {Hern{\'a}ndez}, {Bombrun}, {Ramos-Lerate}, {Steidelm{\"u}ller}, {Bastian},
  {Biermann}, {de Torres}, {Gerlach}, {Geyer}, {Hilger}, {Hobbs}, {Lammers},
  {McMillan}, {Stephenson}, {Casta{\~n}eda}, {Davidson}, {Fabricius},
  {Gracia-Abril}, {Portell}, {Rowell}, {Teyssier}, {Torra}, {Bartolom{\'e}},
  {Clotet}, {Garralda}, {Gonz{\'a}lez-Vidal}, {Torra}, {Abbas}, {Altmann},
  {Anglada Varela}, {Balaguer-N{\'u}{\~n}ez}, {Balog}, {Barache}, {Becciani},
  {Bernet}, {Bertone}, {Bianchi}, {Bouquillon}, {Brown}, {Bucciarelli},
  {Busonero}, {Butkevich}, {Buzzi}, {Cancelliere}, {Carlucci}, {Charlot},
  {Cioni}, {Crosta}, {Crowley}, {del Peloso}, {del Pozo}, {Drimmel}, {Esquej},
  {Fienga}, {Fraile}, {Gai}, {Garcia-Reinaldos}, {Guerra}, {Hambly}, {Hauser},
  {Jan{\ss}en}, {Jordan}, {Kostrzewa-Rutkowska}, {Lattanzi}, {Liao}, {Licata},
  {Lister}, {L{\"o}ffler}, {Marchant}, {Masip}, {Mignard}, {Mints}, {Molina},
  {Mora}, {Morbidelli}, {Murphy}, {Pagani}, {Panuzzo}, {Pe{\~n}alosa Esteller},
  {Poggio}, {Re Fiorentin}, {Riva}, {Sagrist{\`a} Sell{\'e}s}, {Sanchez
  Gimenez}, {Sarasso}, {Sciacca}, {Siddiqui}, {Smart}, {Souami}, {Spagna},
  {Steele}, {Taris}, {Utrilla}, {van Reeven}, \& {Vecchiato}}]{lin21b}
{Lindegren}, L., {Klioner}, S.~A., {Hern{\'a}ndez}, J., {et~al.}
  2021{\natexlab{b}}, \aap, 649, A2, \dodoi{10.1051/0004-6361/202039709}

\bibitem[{{Lucatello} {et~al.}(2015){Lucatello}, {Sollima}, {Gratton},
  {Vesperini}, {D'Orazi}, {Carretta}, \& {Bragaglia}}]{luc15}
{Lucatello}, S., {Sollima}, A., {Gratton}, R., {et~al.} 2015, \aap, 584, A52,
  \dodoi{10.1051/0004-6361/201526957}

\bibitem[{{Luck} \& {Bond}(1981)}]{luc81}
{Luck}, R.~E., \& {Bond}, H.~E. 1981, \apj, 244, 919, \dodoi{10.1086/158767}

\bibitem[{{Luck} \& {Bond}(1985)}]{luc85}
---. 1985, \apj, 292, 559, \dodoi{10.1086/163189}

\bibitem[{{Luri} {et~al.}(2018){Luri}, {Brown}, {Sarro}, {Arenou},
  {Bailer-Jones}, {Castro-Ginard}, {de Bruijne}, {Prusti}, {Babusiaux}, \&
  {Delgado}}]{lur18}
{Luri}, X., {Brown}, A.~G.~A., {Sarro}, L.~M., {et~al.} 2018, \aap, 616, A9,
  \dodoi{10.1051/0004-6361/201832964}

\bibitem[{{Magain}(1985)}]{mag85}
{Magain}, P. 1985, \aap, 146, 95

\bibitem[{{Magain}(1989)}]{mag89}
---. 1989, \aap, 209, 211

\bibitem[{{Majewski} {et~al.}(2017){Majewski}, {Schiavon}, {Frinchaboy},
  {Allende Prieto}, {Barkhouser}, {Bizyaev}, {Blank}, {Brunner}, {Burton},
  {Carrera}, {Chojnowski}, {Cunha}, {Epstein}, {Fitzgerald}, {Garc{\'\i}a
  P{\'e}rez}, {Hearty}, {Henderson}, {Holtzman}, {Johnson}, {Lam}, {Lawler},
  {Maseman}, {M{\'e}sz{\'a}ros}, {Nelson}, {Nguyen}, {Nidever}, {Pinsonneault},
  {Shetrone}, {Smee}, {Smith}, {Stolberg}, {Skrutskie}, {Walker}, {Wilson},
  {Zasowski}, {Anders}, {Basu}, {Beland}, {Blanton}, {Bovy}, {Brownstein},
  {Carlberg}, {Chaplin}, {Chiappini}, {Eisenstein}, {Elsworth}, {Feuillet},
  {Fleming}, {Galbraith-Frew}, {Garc{\'\i}a}, {Garc{\'\i}a-Hern{\'a}ndez},
  {Gillespie}, {Girardi}, {Gunn}, {Hasselquist}, {Hayden}, {Hekker}, {Ivans},
  {Kinemuchi}, {Klaene}, {Mahadevan}, {Mathur}, {Mosser}, {Muna}, {Munn},
  {Nichol}, {O'Connell}, {Parejko}, {Robin}, {Rocha-Pinto}, {Schultheis},
  {Serenelli}, {Shane}, {Silva Aguirre}, {Sobeck}, {Thompson}, {Troup},
  {Weinberg}, \& {Zamora}}]{maj17}
{Majewski}, S.~R., {Schiavon}, R.~P., {Frinchaboy}, P.~M., {et~al.} 2017, \aj,
  154, 94, \dodoi{10.3847/1538-3881/aa784d}

\bibitem[{{Maoz} {et~al.}(2012){Maoz}, {Mannucci}, \& {Brandt}}]{mao12}
{Maoz}, D., {Mannucci}, F., \& {Brandt}, T.~D. 2012, \mnras, 426, 3282,
  \dodoi{10.1111/j.1365-2966.2012.21871.x}

\bibitem[{{Maoz} {et~al.}(2011){Maoz}, {Mannucci}, {Li}, {Filippenko}, {Della
  Valle}, \& {Panagia}}]{mao11}
{Maoz}, D., {Mannucci}, F., {Li}, W., {et~al.} 2011, \mnras, 412, 1508,
  \dodoi{10.1111/j.1365-2966.2010.16808.x}

\bibitem[{{Maoz} {et~al.}(2014){Maoz}, {Mannucci}, \& {Nelemans}}]{mao14}
{Maoz}, D., {Mannucci}, F., \& {Nelemans}, G. 2014, \araa, 52, 107,
  \dodoi{10.1146/annurev-astro-082812-141031}

\bibitem[{{Marcolini} {et~al.}(2006){Marcolini}, {D'Ercole}, {Brighenti}, \&
  {Recchi}}]{mar06}
{Marcolini}, A., {D'Ercole}, A., {Brighenti}, F., \& {Recchi}, S. 2006, \mnras,
  371, 643, \dodoi{10.1111/j.1365-2966.2006.10671.x}

\bibitem[{{Marcolini} {et~al.}(2009){Marcolini}, {Gibson}, {Karakas}, \&
  {S{\'a}nchez-Bl{\'a}zquez}}]{mar09}
{Marcolini}, A., {Gibson}, B.~K., {Karakas}, A.~I., \&
  {S{\'a}nchez-Bl{\'a}zquez}, P. 2009, \mnras, 395, 719,
  \dodoi{10.1111/j.1365-2966.2009.14591.x}

\bibitem[{{Marrese} {et~al.}(2019){Marrese}, {Marinoni}, {Fabrizio}, \&
  {Altavilla}}]{mar19}
{Marrese}, P.~M., {Marinoni}, S., {Fabrizio}, M., \& {Altavilla}, G. 2019,
  \aap, 621, A144, \dodoi{10.1051/0004-6361/201834142}

\bibitem[{{McKinney}(2010)}]{mck10}
{McKinney}, W. 2010, in {P}roceedings of the 9th {P}ython in {S}cience
  {C}onference, ed. {S}t\'efan van~der {W}alt \& {J}arrod {M}illman, 56 -- 61,
  \dodoi{10.25080/Majora-92bf1922-00a}

\bibitem[{{McWilliam} {et~al.}(1995{\natexlab{a}}){McWilliam}, {Preston},
  {Sneden}, \& {Searle}}]{mcw95a}
{McWilliam}, A., {Preston}, G.~W., {Sneden}, C., \& {Searle}, L.
  1995{\natexlab{a}}, \aj, 109, 2757, \dodoi{10.1086/117486}

\bibitem[{{McWilliam} {et~al.}(1995{\natexlab{b}}){McWilliam}, {Preston},
  {Sneden}, \& {Shectman}}]{mcw95b}
{McWilliam}, A., {Preston}, G.~W., {Sneden}, C., \& {Shectman}, S.
  1995{\natexlab{b}}, \aj, 109, 2736, \dodoi{10.1086/117485}

\bibitem[{{Neopane} {et~al.}(2022){Neopane}, {Bhargava}, {Fisher}, {Ferrari},
  {Yoshida}, {Toonen}, \& {Bravo}}]{neo22}
{Neopane}, S., {Bhargava}, K., {Fisher}, R., {et~al.} 2022, \apj, 925, 92,
  \dodoi{10.3847/1538-4357/ac3b52}

\bibitem[{{Nidever} {et~al.}(2015){Nidever}, {Holtzman}, {Allende Prieto},
  {Beland}, {Bender}, {Bizyaev}, {Burton}, {Desphande}, {Fleming}, {Garc{\'\i}a
  P{\'e}rez}, {Hearty}, {Majewski}, {M{\'e}sz{\'a}ros}, {Muna}, {Nguyen},
  {Schiavon}, {Shetrone}, {Skrutskie}, {Sobeck}, \& {Wilson}}]{nid15}
{Nidever}, D.~L., {Holtzman}, J.~A., {Allende Prieto}, C., {et~al.} 2015, \aj,
  150, 173, \dodoi{10.1088/0004-6256/150/6/173}

\bibitem[{{Nidever} {et~al.}(2020){Nidever}, {Hasselquist}, {Hayes}, {Hawkins},
  {Povick}, {Majewski}, {Smith}, {Anguiano}, {Stringfellow}, {Sobeck}, {Cunha},
  {Beers}, {Bestenlehner}, {Cohen}, {Garcia-Hernandez}, {J{\"o}nsson},
  {Nitschelm}, {Shetrone}, {Lacerna}, {Allende Prieto}, {Beaton}, {Dell'Agli},
  {Fern{\'a}ndez-Trincado}, {Feuillet}, {Gallart}, {Hearty}, {Holtzman},
  {Manchado}, {Mu{\~n}oz}, {O'Connell}, \& {Rosado}}]{nid20}
{Nidever}, D.~L., {Hasselquist}, S., {Hayes}, C.~R., {et~al.} 2020, \apj, 895,
  88, \dodoi{10.3847/1538-4357/ab7305}

\bibitem[{{Nomoto} \& {Leung}(2018)}]{nom18}
{Nomoto}, K., \& {Leung}, S.-C. 2018, \ssr, 214, 67,
  \dodoi{10.1007/s11214-018-0499-0}

\bibitem[{{Ochsenbein} {et~al.}(2000){Ochsenbein}, {Bauer}, \&
  {Marcout}}]{och00}
{Ochsenbein}, F., {Bauer}, P., \& {Marcout}, J. 2000, \aaps, 143, 23,
  \dodoi{10.1051/aas:2000169}

\bibitem[{{Ohlmann} {et~al.}(2014){Ohlmann}, {Kromer}, {Fink}, {Pakmor},
  {Seitenzahl}, {Sim}, \& {R{\"o}pke}}]{ohl14}
{Ohlmann}, S.~T., {Kromer}, M., {Fink}, M., {et~al.} 2014, \aap, 572, A57,
  \dodoi{10.1051/0004-6361/201423924}

\bibitem[{{Osorio} {et~al.}(2020){Osorio}, {Allende Prieto}, {Hubeny},
  {M{\'e}sz{\'a}ros}, \& {Shetrone}}]{oso20}
{Osorio}, Y., {Allende Prieto}, C., {Hubeny}, I., {M{\'e}sz{\'a}ros}, S., \&
  {Shetrone}, M. 2020, \aap, 637, A80, \dodoi{10.1051/0004-6361/201937054}

\bibitem[{{pandas Development Team}(2020)}]{reb20}
{pandas Development Team}. 2020, pandas-dev/pandas: Pandas, latest,  Zenodo,
  \dodoi{10.5281/zenodo.3509134}

\bibitem[{{Papish} \& {Perets}(2016)}]{pap16}
{Papish}, O., \& {Perets}, H.~B. 2016, \apj, 822, 19,
  \dodoi{10.3847/0004-637X/822/1/19}

\bibitem[{{Paxton} {et~al.}(2011){Paxton}, {Bildsten}, {Dotter}, {Herwig},
  {Lesaffre}, \& {Timmes}}]{pax11}
{Paxton}, B., {Bildsten}, L., {Dotter}, A., {et~al.} 2011, \apjs, 192, 3,
  \dodoi{10.1088/0067-0049/192/1/3}

\bibitem[{{Paxton} {et~al.}(2013){Paxton}, {Cantiello}, {Arras}, {Bildsten},
  {Brown}, {Dotter}, {Mankovich}, {Montgomery}, {Stello}, {Timmes}, \&
  {Townsend}}]{pax13}
{Paxton}, B., {Cantiello}, M., {Arras}, P., {et~al.} 2013, \apjs, 208, 4,
  \dodoi{10.1088/0067-0049/208/1/4}

\bibitem[{{Paxton} {et~al.}(2015){Paxton}, {Marchant}, {Schwab}, {Bauer},
  {Bildsten}, {Cantiello}, {Dessart}, {Farmer}, {Hu}, {Langer}, {Townsend},
  {Townsley}, \& {Timmes}}]{pax15}
{Paxton}, B., {Marchant}, P., {Schwab}, J., {et~al.} 2015, \apjs, 220, 15,
  \dodoi{10.1088/0067-0049/220/1/15}

\bibitem[{{Paxton} {et~al.}(2018){Paxton}, {Schwab}, {Bauer}, {Bildsten},
  {Blinnikov}, {Duffell}, {Farmer}, {Goldberg}, {Marchant}, {Sorokina},
  {Thoul}, {Townsend}, \& {Timmes}}]{pax18}
{Paxton}, B., {Schwab}, J., {Bauer}, E.~B., {et~al.} 2018, \apjs, 234, 34,
  \dodoi{10.3847/1538-4365/aaa5a8}

\bibitem[{{Peterson}(1981)}]{pet81}
{Peterson}, R.~C. 1981, \apj, 244, 989, \dodoi{10.1086/158771}

\bibitem[{{Pilachowski} {et~al.}(1983){Pilachowski}, {Sneden}, \&
  {Wallerstein}}]{pil83}
{Pilachowski}, C.~A., {Sneden}, C., \& {Wallerstein}, G. 1983, \apjs, 52, 241,
  \dodoi{10.1086/190867}

\bibitem[{{Plez}(2012)}]{ple12}
{Plez}, B. 2012, {Turbospectrum: Code for spectral synthesis}, Astrophysics
  Source Code Library, record ascl:1205.004.
\newblock \doeprint{1205.004}

\bibitem[{{Pomp{\'e}ia} {et~al.}(2008){Pomp{\'e}ia}, {Hill}, {Spite}, {Cole},
  {Primas}, {Romaniello}, {Pasquini}, {Cioni}, \& {Smecker Hane}}]{pom08}
{Pomp{\'e}ia}, L., {Hill}, V., {Spite}, M., {et~al.} 2008, \aap, 480, 379,
  \dodoi{10.1051/0004-6361:20064854}

\bibitem[{{R Core Team}(2023)}]{r23}
{R Core Team}. 2023, R: A Language and Environment for Statistical Computing, R
  Foundation for Statistical Computing, Vienna, Austria.
\newblock \url{https://www.R-project.org/}

\bibitem[{{Reggiani} {et~al.}(2023){Reggiani}, {Schlaufman}, \&
  {Casey}}]{reg23}
{Reggiani}, H., {Schlaufman}, K.~C., \& {Casey}, A.~R. 2023, arXiv e-prints,
  arXiv:2303.16357, \dodoi{10.48550/arXiv.2303.16357}

\bibitem[{{Reggiani} {et~al.}(2021){Reggiani}, {Schlaufman}, {Casey}, {Simon},
  \& {Ji}}]{reg21}
{Reggiani}, H., {Schlaufman}, K.~C., {Casey}, A.~R., {Simon}, J.~D., \& {Ji},
  A.~P. 2021, \aj, 162, 229, \dodoi{10.3847/1538-3881/ac1f9a}

\bibitem[{{Riello} {et~al.}(2018){Riello}, {De Angeli}, {Evans}, {Busso},
  {Hambly}, {Davidson}, {Burgess}, {Montegriffo}, {Osborne}, {Kewley},
  {Carrasco}, {Fabricius}, {Jordi}, {Cacciari}, {van Leeuwen}, \&
  {Holland}}]{rie18}
{Riello}, M., {De Angeli}, F., {Evans}, D.~W., {et~al.} 2018, \aap, 616, A3,
  \dodoi{10.1051/0004-6361/201832712}

\bibitem[{{Riello} {et~al.}(2021){Riello}, {De Angeli}, {Evans}, {Montegriffo},
  {Carrasco}, {Busso}, {Palaversa}, {Burgess}, {Diener}, {Davidson}, {Rowell},
  {Fabricius}, {Jordi}, {Bellazzini}, {Pancino}, {Harrison}, {Cacciari}, {van
  Leeuwen}, {Hambly}, {Hodgkin}, {Osborne}, {Altavilla}, {Barstow}, {Brown},
  {Castellani}, {Cowell}, {De Luise}, {Gilmore}, {Giuffrida}, {Hidalgo},
  {Holland}, {Marinoni}, {Pagani}, {Piersimoni}, {Pulone}, {Ragaini}, {Rainer},
  {Richards}, {Sanna}, {Walton}, {Weiler}, \& {Yoldas}}]{rie21}
---. 2021, \aap, 649, A3, \dodoi{10.1051/0004-6361/202039587}

\bibitem[{{Rowell} {et~al.}(2021){Rowell}, {Davidson}, {Lindegren}, {van
  Leeuwen}, {Casta{\~n}eda}, {Fabricius}, {Bastian}, {Hambly}, {Hern{\'a}ndez},
  {Bombrun}, {Evans}, {De Angeli}, {Riello}, {Busonero}, {Crowley}, {Mora},
  {Lammers}, {Gracia}, {Portell}, {Biermann}, \& {Brown}}]{row21}
{Rowell}, N., {Davidson}, M., {Lindegren}, L., {et~al.} 2021, \aap, 649, A11,
  \dodoi{10.1051/0004-6361/202039448}

\bibitem[{{Ryan} {et~al.}(1991){Ryan}, {Norris}, \& {Bessell}}]{rya91}
{Ryan}, S.~G., {Norris}, J.~E., \& {Bessell}, M.~S. 1991, \aj, 102, 303,
  \dodoi{10.1086/115878}

\bibitem[{{Sakari} {et~al.}(2018){Sakari}, {Placco}, {Farrell}, {Roederer},
  {Wallerstein}, {Beers}, {Ezzeddine}, {Frebel}, {Hansen}, {Holmbeck},
  {Sneden}, {Cowan}, {Venn}, {Davis}, {Matijevi{\v{c}}}, {Wyse},
  {Bland-Hawthorn}, {Chiappini}, {Freeman}, {Gibson}, {Grebel}, {Helmi},
  {Kordopatis}, {Kunder}, {Navarro}, {Reid}, {Seabroke}, {Steinmetz}, \&
  {Watson}}]{sak18}
{Sakari}, C.~M., {Placco}, V.~M., {Farrell}, E.~M., {et~al.} 2018, \apj, 868,
  110, \dodoi{10.3847/1538-4357/aae9df}

\bibitem[{{Salgado} {et~al.}(2017){Salgado}, {Gonz{\'a}lez-N{\'u}{\~n}ez},
  {Guti{\'e}rrez-S{\'a}nchez}, {Segovia}, {Dur{\'a}n}, {Hern{\'a}ndez}, \&
  {Arviset}}]{sal17}
{Salgado}, J., {Gonz{\'a}lez-N{\'u}{\~n}ez}, J., {Guti{\'e}rrez-S{\'a}nchez},
  R., {et~al.} 2017, Astronomy and Computing, 21, 22,
  \dodoi{10.1016/j.ascom.2017.08.002}

\bibitem[{{S{\'a}nchez-Bl{\'a}zquez} {et~al.}(2012){S{\'a}nchez-Bl{\'a}zquez},
  {Marcolini}, {Gibson}, {Karakas}, {Pilkington}, \& {Calura}}]{san12}
{S{\'a}nchez-Bl{\'a}zquez}, P., {Marcolini}, A., {Gibson}, B.~K., {et~al.}
  2012, \mnras, 419, 1376, \dodoi{10.1111/j.1365-2966.2011.19793.x}

\bibitem[{{Santana} {et~al.}(2021){Santana}, {Beaton}, {Covey}, {O'Connell},
  {Longa-Pe{\~n}a}, {Cohen}, {Fern{\'a}ndez-Trincado}, {Hayes}, {Zasowski},
  {Sobeck}, {Majewski}, {Chojnowski}, {De Lee}, {Oelkers}, {Stringfellow},
  {Almeida}, {Anguiano}, {Donor}, {Frinchaboy}, {Hasselquist}, {Johnson},
  {Kollmeier}, {Nidever}, {Price-Whelan}, {Rojas-Arriagada}, {Schultheis},
  {Shetrone}, {Simon}, {Aerts}, {Borissova}, {Drout}, {Geisler}, {Law},
  {Medina}, {Minniti}, {Monachesi}, {Mu{\~n}oz}, {Poleski}, {Roman-Lopes},
  {Schlaufman}, {Stutz}, {Teske}, {Tkachenko}, {Van Saders}, {Weinberger}, \&
  {Zoccali}}]{san21}
{Santana}, F.~A., {Beaton}, R.~L., {Covey}, K.~R., {et~al.} 2021, \aj, 162,
  303, \dodoi{10.3847/1538-3881/ac2cbc}

\bibitem[{{Schlaufman}(2014)}]{sch14}
{Schlaufman}, K.~C. 2014, \apj, 790, 91, \dodoi{10.1088/0004-637X/790/2/91}

\bibitem[{{Seitenzahl} {et~al.}(2013){Seitenzahl}, {Ciaraldi-Schoolmann},
  {R{\"o}pke}, {Fink}, {Hillebrandt}, {Kromer}, {Pakmor}, {Ruiter}, {Sim}, \&
  {Taubenberger}}]{sei13}
{Seitenzahl}, I.~R., {Ciaraldi-Schoolmann}, F., {R{\"o}pke}, F.~K., {et~al.}
  2013, \mnras, 429, 1156, \dodoi{10.1093/mnras/sts402}

\bibitem[{{Seitenzahl} {et~al.}(2016){Seitenzahl}, {Kromer}, {Ohlmann},
  {Ciaraldi-Schoolmann}, {Marquardt}, {Fink}, {Hillebrandt}, {Pakmor},
  {R{\"o}pke}, {Ruiter}, {Sim}, \& {Taubenberger}}]{sei16}
{Seitenzahl}, I.~R., {Kromer}, M., {Ohlmann}, S.~T., {et~al.} 2016, \aap, 592,
  A57, \dodoi{10.1051/0004-6361/201527251}

\bibitem[{{Shetrone} {et~al.}(2003){Shetrone}, {Venn}, {Tolstoy}, {Primas},
  {Hill}, \& {Kaufer}}]{she03}
{Shetrone}, M., {Venn}, K.~A., {Tolstoy}, E., {et~al.} 2003, \aj, 125, 684,
  \dodoi{10.1086/345966}

\bibitem[{{Shetrone} {et~al.}(2015){Shetrone}, {Bizyaev}, {Lawler}, {Allende
  Prieto}, {Johnson}, {Smith}, {Cunha}, {Holtzman}, {Garc{\'\i}a P{\'e}rez},
  {M{\'e}sz{\'a}ros}, {Sobeck}, {Zamora}, {Garc{\'\i}a-Hern{\'a}ndez}, {Souto},
  {Chojnowski}, {Koesterke}, {Majewski}, \& {Zasowski}}]{she15}
{Shetrone}, M., {Bizyaev}, D., {Lawler}, J.~E., {et~al.} 2015, \apjs, 221, 24,
  \dodoi{10.1088/0067-0049/221/2/24}

\bibitem[{{Shetrone} {et~al.}(2001){Shetrone}, {C{\^o}t{\'e}}, \&
  {Sargent}}]{she01}
{Shetrone}, M.~D., {C{\^o}t{\'e}}, P., \& {Sargent}, W.~L.~W. 2001, \apj, 548,
  592, \dodoi{10.1086/319022}

\bibitem[{{Skrutskie} {et~al.}(2006){Skrutskie}, {Cutri}, {Stiening},
  {Weinberg}, {Schneider}, {Carpenter}, {Beichman}, {Capps}, {Chester},
  {Elias}, {Huchra}, {Liebert}, {Lonsdale}, {Monet}, {Price}, {Seitzer},
  {Jarrett}, {Kirkpatrick}, {Gizis}, {Howard}, {Evans}, {Fowler}, {Fullmer},
  {Hurt}, {Light}, {Kopan}, {Marsh}, {McCallon}, {Tam}, {Van Dyk}, \&
  {Wheelock}}]{skr06}
{Skrutskie}, M.~F., {Cutri}, R.~M., {Stiening}, R., {et~al.} 2006, \aj, 131,
  1163, \dodoi{10.1086/498708}

\bibitem[{{Smith} {et~al.}(2013){Smith}, {Cunha}, {Shetrone}, {Meszaros},
  {Allende Prieto}, {Bizyaev}, {Garc{\'\i}a P{\'e}rez}, {Majewski}, {Schiavon},
  {Holtzman}, \& {Johnson}}]{smi13}
{Smith}, V.~V., {Cunha}, K., {Shetrone}, M.~D., {et~al.} 2013, \apj, 765, 16,
  \dodoi{10.1088/0004-637X/765/1/16}

\bibitem[{{Smith} {et~al.}(2021){Smith}, {Bizyaev}, {Cunha}, {Shetrone},
  {Souto}, {Allende Prieto}, {Masseron}, {M{\'e}sz{\'a}ros}, {J{\"o}nsson},
  {Hasselquist}, {Osorio}, {Garc{\'\i}a-Hern{\'a}ndez}, {Plez}, {Beaton},
  {Holtzman}, {Majewski}, {Stringfellow}, \& {Sobeck}}]{smi21}
{Smith}, V.~V., {Bizyaev}, D., {Cunha}, K., {et~al.} 2021, \aj, 161, 254,
  \dodoi{10.3847/1538-3881/abefdc}

\bibitem[{{Sneden} {et~al.}(1992){Sneden}, {Kraft}, {Prosser}, \&
  {Langer}}]{sne92}
{Sneden}, C., {Kraft}, R.~P., {Prosser}, C.~F., \& {Langer}, G.~E. 1992, \aj,
  104, 2121, \dodoi{10.1086/116388}

\bibitem[{{Sukhbold} {et~al.}(2016){Sukhbold}, {Ertl}, {Woosley}, {Brown}, \&
  {Janka}}]{suk16}
{Sukhbold}, T., {Ertl}, T., {Woosley}, S.~E., {Brown}, J.~M., \& {Janka}, H.~T.
  2016, \apj, 821, 38, \dodoi{10.3847/0004-637X/821/1/38}

\bibitem[{{Tolstoy} {et~al.}(2003){Tolstoy}, {Venn}, {Shetrone}, {Primas},
  {Hill}, {Kaufer}, \& {Szeifert}}]{tol03}
{Tolstoy}, E., {Venn}, K.~A., {Shetrone}, M., {et~al.} 2003, \aj, 125, 707,
  \dodoi{10.1086/345967}

\bibitem[{{Torra} {et~al.}(2021){Torra}, {Casta{\~n}eda}, {Fabricius},
  {Lindegren}, {Clotet}, {Gonz{\'a}lez-Vidal}, {Bartolom{\'e}}, {Bastian},
  {Bernet}, {Biermann}, {Garralda}, {G{\'u}rpide}, {Lammers}, {Portell}, \&
  {Torra}}]{tor21}
{Torra}, F., {Casta{\~n}eda}, J., {Fabricius}, C., {et~al.} 2021, \aap, 649,
  A10, \dodoi{10.1051/0004-6361/202039637}

\bibitem[{{Totani} {et~al.}(2008){Totani}, {Morokuma}, {Oda}, {Doi}, \&
  {Yasuda}}]{tot08}
{Totani}, T., {Morokuma}, T., {Oda}, T., {Doi}, M., \& {Yasuda}, N. 2008,
  \pasj, 60, 1327, \dodoi{10.1093/pasj/60.6.1327}

\bibitem[{{Van der Swaelmen} {et~al.}(2013){Van der Swaelmen}, {Hill},
  {Primas}, \& {Cole}}]{van13}
{Van der Swaelmen}, M., {Hill}, V., {Primas}, F., \& {Cole}, A.~A. 2013, \aap,
  560, A44, \dodoi{10.1051/0004-6361/201321109}

\bibitem[{{Vasiliev} \& {Baumgardt}(2021)}]{vas21}
{Vasiliev}, E., \& {Baumgardt}, H. 2021, \mnras, 505, 5978,
  \dodoi{10.1093/mnras/stab1475}

\bibitem[{{Virtanen} {et~al.}(2020){Virtanen}, {Gommers}, {Oliphant},
  {Haberland}, {Reddy}, {Cournapeau}, {Burovski}, {Peterson}, {Weckesser},
  {Bright}, {van der Walt}, {Brett}, {Wilson}, {Millman}, {Mayorov}, {Nelson},
  {Jones}, {Kern}, {Larson}, {Carey}, {Polat}, {Feng}, {Moore}, {VanderPlas},
  {Laxalde}, {Perktold}, {Cimrman}, {Henriksen}, {Quintero}, {Harris},
  {Archibald}, {Ribeiro}, {Pedregosa}, {van Mulbregt}, \& {SciPy 1. 0
  Contributors}}]{vir20}
{Virtanen}, P., {Gommers}, R., {Oliphant}, T.~E., {et~al.} 2020, Nature
  Methods, 17, 261, \dodoi{10.1038/s41592-019-0686-2}

\bibitem[{{Wallerstein}(1962)}]{wal62}
{Wallerstein}, G. 1962, \apjs, 6, 407, \dodoi{10.1086/190067}

\bibitem[{{Wenger} {et~al.}(2000){Wenger}, {Ochsenbein}, {Egret}, {Dubois},
  {Bonnarel}, {Borde}, {Genova}, {Jasniewicz}, {Lalo{\"e}}, {Lesteven}, \&
  {Monier}}]{wen00}
{Wenger}, M., {Ochsenbein}, F., {Egret}, D., {et~al.} 2000, \aaps, 143, 9,
  \dodoi{10.1051/aas:2000332}

\bibitem[{{Wilson} {et~al.}(2019){Wilson}, {Hearty}, {Skrutskie}, {Majewski},
  {Holtzman}, {Eisenstein}, {Gunn}, {Blank}, {Henderson}, {Smee}, {Nelson},
  {Nidever}, {Arns}, {Barkhouser}, {Barr}, {Beland}, {Bershady}, {Blanton},
  {Brunner}, {Burton}, {Carey}, {Carr}, {Colque}, {Crane}, {Damke}, {Davidson},
  {Dean}, {Di Mille}, {Don}, {Ebelke}, {Evans}, {Fitzgerald}, {Gillespie},
  {Hall}, {Harding}, {Harding}, {Hammond}, {Hancock}, {Harrison}, {Hope},
  {Horne}, {Karakla}, {Lam}, {Leger}, {MacDonald}, {Maseman}, {Matsunari},
  {Melton}, {Mitcheltree}, {O'Brien}, {O'Connell}, {Patten}, {Richardson},
  {Rieke}, {Rieke}, {Roman-Lopes}, {Schiavon}, {Sobeck}, {Stolberg}, {Stoll},
  {Tembe}, {Trujillo}, {Uomoto}, {Vernieri}, {Walker}, {Weinberg}, {Young},
  {Anthony-Brumfield}, {Bizyaev}, {Breslauer}, {De Lee}, {Downey}, {Halverson},
  {Huehnerhoff}, {Klaene}, {Leon}, {Long}, {Mahadevan}, {Malanushenko},
  {Nguyen}, {Owen}, {S{\'a}nchez-Gallego}, {Sayres}, {Shane}, {Shectman},
  {Shetrone}, {Skinner}, {Stauffer}, \& {Zhao}}]{wil19}
{Wilson}, J.~C., {Hearty}, F.~R., {Skrutskie}, M.~F., {et~al.} 2019, \pasp,
  131, 055001, \dodoi{10.1088/1538-3873/ab0075}

\bibitem[{Xing {et~al.}(2023)Xing, Zhao, Liu, Heger, Han, Aoki, Chen, Ishigaki,
  Li, \& Zhao}]{xin23}
Xing, Q.-F., Zhao, G., Liu, Z.-W., {et~al.} 2023, Nature, 618, 712,
  \dodoi{10.1038/s41586-023-06028-1}

\bibitem[{{Zasowski} {et~al.}(2013){Zasowski}, {Johnson}, {Frinchaboy},
  {Majewski}, {Nidever}, {Rocha Pinto}, {Girardi}, {Andrews}, {Chojnowski},
  {Cudworth}, {Jackson}, {Munn}, {Skrutskie}, {Beaton}, {Blake}, {Covey},
  {Deshpande}, {Epstein}, {Fabbian}, {Fleming}, {Garcia Hernandez}, {Herrero},
  {Mahadevan}, {M{\'e}sz{\'a}ros}, {Schultheis}, {Sellgren}, {Terrien}, {van
  Saders}, {Allende Prieto}, {Bizyaev}, {Burton}, {Cunha}, {da Costa},
  {Hasselquist}, {Hearty}, {Holtzman}, {Garc{\'\i}a P{\'e}rez}, {Maia},
  {O'Connell}, {O'Donnell}, {Pinsonneault}, {Santiago}, {Schiavon}, {Shetrone},
  {Smith}, \& {Wilson}}]{zas13}
{Zasowski}, G., {Johnson}, J.~A., {Frinchaboy}, P.~M., {et~al.} 2013, \aj, 146,
  81, \dodoi{10.1088/0004-6256/146/4/81}

\bibitem[{{Zasowski} {et~al.}(2017){Zasowski}, {Cohen}, {Chojnowski},
  {Santana}, {Oelkers}, {Andrews}, {Beaton}, {Bender}, {Bird}, {Bovy},
  {Carlberg}, {Covey}, {Cunha}, {Dell'Agli}, {Fleming}, {Frinchaboy},
  {Garc{\'\i}a-Hern{\'a}ndez}, {Harding}, {Holtzman}, {Johnson}, {Kollmeier},
  {Majewski}, {M{\'e}sz{\'a}ros}, {Munn}, {Mu{\~n}oz}, {Ness}, {Nidever},
  {Poleski}, {Rom{\'a}n-Z{\'u}{\~n}iga}, {Shetrone}, {Simon}, {Smith},
  {Sobeck}, {Stringfellow}, {Szigeti{\'a}ros}, {Tayar}, \& {Troup}}]{zas17}
{Zasowski}, G., {Cohen}, R.~E., {Chojnowski}, S.~D., {et~al.} 2017, \aj, 154,
  198, \dodoi{10.3847/1538-3881/aa8df9}

\end{thebibliography}
\bibliographystyle{aasjournal}

\listofchanges
\end{document}